\documentclass[journal]{IEEEtran}

\ifCLASSINFOpdf
\else
\fi
\usepackage{booktabs}
\usepackage{amssymb}
\usepackage{amsthm}
\usepackage{graphicx}
\usepackage{subfigure}
\usepackage{mathptmx}      
\usepackage{amsmath}
\usepackage{latexsym}
\usepackage{bm}
\usepackage{balance}
\usepackage{caption}

\newtheorem{thm}{Theorem}

\usepackage{booktabs}
\usepackage[numbers,sort&compress]{natbib}
\usepackage{makecell,multirow,diagbox}
\newcommand{\tabincell}[2]{\begin{tabular}{@{}#1@{}}#2\end{tabular}}
\usepackage{threeparttable}
\usepackage[colorlinks,urlcolor=blue]{hyperref} 

\begin{document}
%
\title{AutoBCS: Block-based Image Compressive Sensing with Data-driven Acquisition and Non-iterative Reconstruction}
%
%
%

\author{Hongping Gan,~\IEEEmembership{Member,~IEEE}, Yang Gao, Chunyi Liu, Haiwei Chen, \\ Tao~Zhang,~\IEEEmembership{Member,~IEEE}, and Feng Liu\\

\thanks{This work was supported in part by the Fundamental Research Funds for the Central Universities under Grant G2020KY05110, in part by the Basic Research Programs of Taicang under Grant TC2020JC07, in part by the China Postdoctoral Science Foundation under Grant 2020M680562, in part by the National Key R$\&$D Program of China under Grant 2017YFB0502700, in part by the National Natural Science Foundation of China under Grant 61372069.}
\thanks{Hongping Gan is with the School of Software, Northwestern Polytechnical University, Xi'an 710072, China (e-mail: ganhongping@nwpu.edu.cn).}
\thanks{Yang Gao, Chunyi Liu, Haiwei Chen and Feng Liu are with the School of Information Technology and Electrical Engineering, The University of Queensland, Brisbane, QLD 4072, Australia (e-mail: \{yang.gao; chunyi.liu; haiwei.chen; feng\}@uq.edu.au).}
\thanks{Tao Zhang is with the Department of Electronic Engineering, Tsinghua University, Beijing 100084, China (e-mail: zhangtao8902@mail.tsinghua.edu.cn).}


}

%
%

\markboth{}%
{Shell \MakeLowercase{\textit{et al.}}: Bare Demo of IEEEtran.cls for IEEE Journals}
%



\maketitle

\begin{abstract}
Block compressive sensing is a well-known signal acquisition and reconstruction paradigm with widespread application prospects in science, engineering and cybernetic systems. However, state-of-the-art block-based image compressive sensing (BCS) methods generally suffer from two issues. The sparsifying domain and the sensing matrices widely used for image acquisition are not data-driven, and thus both the features of the image and the relationships among subblock images are ignored. Moreover, doing so requires addressing high-dimensional optimization problems with extensive computational complexity for image reconstruction. In this paper, we provide a deep learning strategy for BCS, called AutoBCS, which takes the prior knowledge of images into account in the acquisition step and establishes a subsequent reconstruction model for performing fast image reconstruction with a low computational cost. More precisely, we present a learning-based sensing matrix (LSM) derived from training data to accomplish image acquisition, thereby capturing and preserving more image characteristics than those captured by existing methods. In particular, the generated LSM is proven to satisfy the theoretical requirements of compressive sensing, such as the so-called restricted isometry property. Additionally, we build a noniterative reconstruction network, which provides an end-to-end BCS reconstruction framework to eliminate blocking artifacts and maximize image reconstruction accuracy, in our AutoBCS architecture. Furthermore, we investigate comprehensive comparison studies with both traditional BCS approaches and newly developed deep learning methods. Compared with these approaches, our AutoBCS framework can not only provide superior performance in terms of image quality metrics (SSIM and PSNR) and visual perception, but also automatically benefit reconstruction speed. The code is available on \href{https://github.com/YangGaoUQ/AutoBCS/}{AutoBCS}\footnote{Please check the webpage {https://github.com/YangGaoUQ/AutoBCS/.}}.
\end{abstract}

%
\IEEEpeerreviewmaketitle

\section{INTRODUCTION}
\label{sec:1}

Compressive sensing (CS) has renewed explosive interest in essential sampling techniques with the goal of acquiring and reconstructing signals at sub-Nyquist sampling rates. CS automatically enables \emph{hardware data compression} of the signals of interest $\bm{x} \in \mathbb{R}^{n}$ during the sampling process, and this provides significant potential for increasing the energy efficiency of sensors in modern signal processing applications. Under the guidance of a series of landmark works by Cand\`{e}s \cite{Candes1} and Donoho \cite{donoho2006compressed}, a large number of CS-based applications have emerged in recent years, such as quantum state tomography \cite{Zhang8354810}, data security \cite{Zhang7492261,Zhang8516287}, multilayer networks \cite{mei2017compressive}, and transcriptomic profiling \cite{Zhang8906045}.

Mathematically, CS theory states that $\bm{x}$ can be exactly reconstructed from a set of non-adaptive measurements formed by ${\bm{y}=\bm{B} {\cdot} \bm{x}}$, where $\bm{B}$ is of size $ {m\times n} (m<<n) $, assuming that $\bm{x}$ satisfies the sparsity property and the sensing matrix (sampling patterns) $\bm{B}$ meets certain structural conditions. Note that the number of measurements, $m$, is on the order of the information-theoretic dimension instead of that of the linear-algebraic ambient dimension of $\bm{x}$ \cite{eldar2012compressed,Liu6880351,Zhou7900408}. Obviously, the sparsity property of $\bm{x}$, the sensing matrix $\bm{B}$, and nonlinear reconstruction are three ingredients for perfect reconstruction in CS theory.

\textbf{Sparsity property.} Signals of interest can always be well represented as a linear combination with just a few entries from a certain dictionary or basis. Let us first focus on the concept of a sparse signal. A signal $\bm{x}$ is regarded as $s$-sparse when $\|\bm{x}\|_{0} \leq s$. Let
\begin{equation}
\Sigma_s =\{\bm{x}:\|\bm{x}\|_0 \leq s\},\label{eq:1}
\end{equation}
be an array of all $s$-sparse signals. In general, we always deal with the fact that $\bm{x}$ is not itself sparse, but it allows for a well-approximated sparse signal in a sparsifying basis $\bm{\Psi} \in \mathbb{R}^{n}$. If so, we can write $\bm{x}$ as $\bm{x}=\bm{\Psi} \cdot \bm{v}$ with $\bm{v}$ being $s$-sparse. Originally, typical choices of the transform basis for sparse image regularization include the discrete cosine transform, wavelet domain, etc. However, these conventional sparse regularizations cannot seize the higher-order dependency of the coefficient vector $\bm{v}$. To capture such dependencies, various specialized and sophisticated regularizations have been exploited for CS with images, most remarkably, attribute correlation learning \cite{Lu7422756,Wu7582406,Lin7875418}, group/structured sparsity \cite{lai2016image}, Bayesian/model-based sparsity \cite{Baraniuk2010}, low-rank regularization \cite{ravishankar2017low}, and nonlocal sparsity \cite{dong2014compressive}. These sparse regularizations can develop an interpretable sparsity prior of $\bm {x}$, resulting in the corresponding reconstruction approaches often providing more accurate and efficient reconstruction results than other techniques. For optimal CS recovery, unfortunately, the aforementioned sparsity regularization methods are either not learned in a data-driven way, or they are not predefined in CS-based image applications.

\textbf{Sensing matrix.} The sensing matrix $\bm{B}$ should satisfy certain structural properties to seize and preserve the salient information of $\bm{x}$ during linear dimensionality reduction: $\mathbb{R}^{n}\rightarrow\mathbb{R}^{m}$. An elegant property of $\bm{B}$ that guarantees such signal acquisition is called the restricted isometry property (RIP) \cite{candes2008restricted}, which holds if $\bm{B}$ approaches $\Sigma_s$ as an approximate isometry. More formally, a matrix $\bm{B}$ satisfies the $(s,\delta_s)$-RIP if
\begin{eqnarray}
\forall \bm{x} \in \Sigma_s,  \; \quad (1-{\delta})   \leq \frac  {{\|{{\bm{B}} \bm{x}}\|}_{2}^2} { {\|\bm{x}\|}_{2}^2}  \leq (1+{\delta}),  \label{eq:2}
\end{eqnarray}
where the smallest constant $\delta_s \leq \delta$ obeying Eq.~\eqref{eq:2} is referred to as the restricted isometry constant. Other famous properties for selecting $\bm{B}$ include the coherence \cite{elad2010sparse}, spark \cite{eldar2012compressed}, and null space properties \cite{cohen2009compressed}. Based on these guiding properties, numerous insightful sensing matrices have been introduced for CS applications \cite{li2014deterministic,gan2018construction,gan2019chaotic}. In particular, it is widely believed that random matrices (e.g., a Gaussian sensing matrix) can allow us to perfectly reconstruct $\bm{x}$ from only $\mathcal{O} (s {\log (n/s) })$ measurements in polynomial time through different reconstruction algorithms \cite{eldar2012compressed}. However, the sensing matrices used in practical CS applications are often signal independent, so they discard some prior signal information.

\textbf{Nonlinear reconstruction.} With the obtained measurements $\bm{y}$ and the transform sparsity of $\bm{x}$, it is a well-established fact that we can reconstruct the original signal $\bm{x}$ via the following optimization problem:
\begin{equation}
\tilde{\bm{x}} =\arg \min_{\bm{x}} \frac{1}{2} \|{{\bm{B}} \bm{x}} - \bm{y}\|_2^2 + \lambda  \|\bm{\Psi x}\|_1 , \label{eq:3}
\end{equation}
where $\lambda$ denotes the regularization parameter. We can refer to this procedure as nonlinear reconstruction mapping $\tilde{\bm{x}}=f(\bm{y})$. An intuitive way to solve Eq.~\eqref{eq:3} is to utilize a convex programming method. However, such a method often suffers from high computational complexity when handling large signals, such as images. To avoid this, a variety of low-cost iterative approaches have been developed, including alternating iterative maximization-minimization \cite{fan2019compressed}, the variational alternating direction method of multipliers (vADMM) \cite{zhang2020signal}, and iterative reweighted least squares (IRLS) \cite{chen2018fast}. Although these algorithms benefit from guaranteed convergence and theoretical analysis, the involved parameters (e.g., penalty parameters and step sizes) are always handcrafted in the solvers. Moreover, they usually require hundreds of iterations to converge to a (sub)optimal solution, leading to the whole process being very time consuming, possibly taking a few minutes.

In summary, the aforementioned three issues are the core issues of CS. From a theoretical perspective, the general-purpose CS approach is always effective. However, practitioners in most engineering applications often face industrial bottlenecks, such as big data problems, high memory requirements, and slow reconstruction speeds. Many efficient strategies have been exploited to transfer CS from theory to industry, including block CS \cite{monika2020adaptive}, infinite dimensional CS \cite{adcock2018infinite}, and quantized CS \cite{huynh2020fast}.
\begin{figure*}[ht]
\center
\includegraphics [width=0.88 \textwidth,height=0.34 \textwidth]{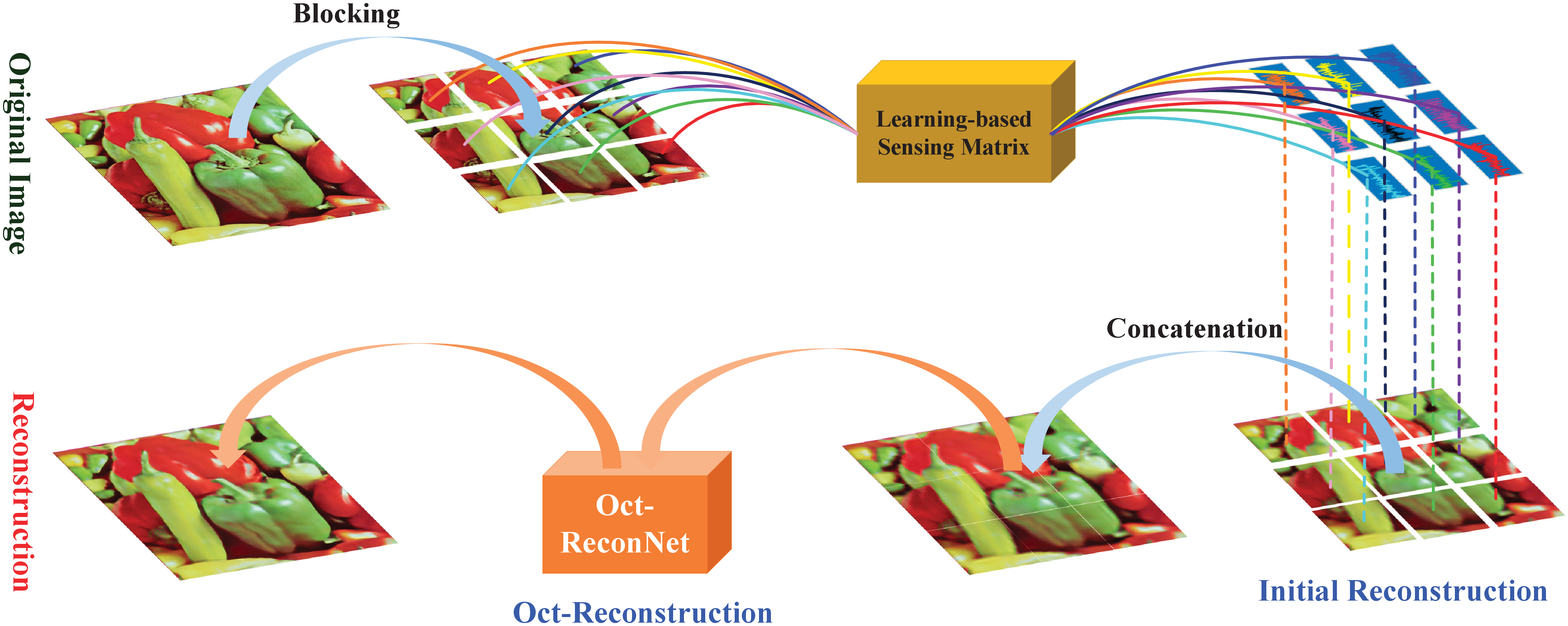}
\caption{Schematic representation of our proposed AutoBCS architecture. AutoBCS replaces the traditional BCS approach with a unified image acquisition and reconstruction framework.}
\label{Fig:1}
\end{figure*}

\subsection{Motivation}
\label{sec:1.1}
Over the past decade, block CS has drawn particular interest because it has exhibited widespread application prospects in science and engineering, such as sparse subspace clustering \cite{cevher2009recovery}, spectrum sensing \cite{polo2009compressive}, and multiple measurement vectors \cite{ji2008multitask}. In particular, such a block-based CS framework has been successfully adapted to image acquisition systems to relieve the memory and energy consumption burdens of sensors and reconstruction algorithms \cite{gan2007block}. As a consequence, block-based image compressive sensing (BCS) is extraordinarily desirable for low-power imaging devices (e.g., wireless visual sensor networks) because of their limited computational capabilities. Compared with the traditional general-purpose CS approach for images, BCS usually benefits from the following aspects. First, block-based measurements are naturally suitable for real-time, energy-limited imaging applications since they avoids dealing with high-dimensional data. Second, we can accelerate the image reconstruction process as each subblock image is independently handled. Third, the sensing operator only takes a small amount of memory and corresponds to feasible hardware architectures due to its structured construction.

The initial BCS framework introduced by Gan et al. \cite{gan2007block} contains two separate phases: sampling and reconstruction. In the sampling phase, the original signal/image of interest is divided into nonoverlapping subblock images, and then each subblock is independently measured via the same sensing operator. In other words, the so-called BCS procedure accomplishes image acquisition in a block-by-block style. We can consider the equivalent full sensing matrix in BCS as a block diagonal matrix whose subblocks are identical copies of the utilized sensing operator. During the reconstruction phase, linear estimation coupled with two-stage iterative hard thresholding is devoted to reconstructing the original image. Following this baseline framework, some sophistical reconstruction algorithms have been proposed to further improve the performance of BCS by exploiting either extra optimization criteria \cite{mun2009block,fowler2011multiscale} or image priors \cite{zhang2012image,zhang2014group}; these methods include smoothed projected Landweber reconstruction (SPL) \cite{mun2009block,chen2011compressed}, collaborative sparsity (RCoS) \cite{zhang2012image}, group-based sparse representation (GBsR) \cite{zhang2014group}, and wavelet packet thresholding (WaPT) \cite{zhao2019block}.

Although the aforementioned algorithms have exhibited promising performance and are beneficial with theoretical analysis, they still encounter the challenge of having to adjust their parameters because they are essentially iterative algorithms for solving Eq.~\eqref{eq:3}. Moreover, the transform domain and sensing operator in BCS are handcrafted (not data-driven) and therefore do not adequately make use of the prior knowledge of images, as described before. To overcome these problems, we are unsurprisingly motivated to seek a data-driven BCS strategy, that is, a method for learning BCS from a set of training data. To this end, one should center on the following questions:
\begin{itemize}
  \item Can we design a learning-based image acquisition way without handcrafting both the sparsifying domain and the sensing operator?
  \item With data-dependent acquisition, is there any possibility for us to customize noniterative reconstruction algorithms for BCS?
  \item How does the performance of the data-driven BCS framework compare to those of the state-of-the-art algorithms?
\end{itemize}

These questions naturally guide us to incorporate the concept of deep learning (DL) in BCS. Roughly speaking, DL is a framework for automating feature learning and extraction, and it has been broadly used in computer vision tasks. Inspired by this technique, different attempts have been exploited to adopt the DL concept for CS imaging. First, improvements to iterative algorithms using DL techniques, such as ADMM-CSnet \cite{yang2018admm} and ISTA-net \cite{zhang2018ista}, have been introduced for image reconstruction; however, these frameworks still employ the traditional sensing operator, and thus, the data acquisition process is signal independent. Additionally, block-based reconstruction networks for BCS, such as ReconNet \cite{kulkarni2016reconnet} and DR$^2$-net \cite{yao2019dr2}, have been developed; nevertheless, these networks only utilize intra-block information to recover a subblock, thereby yielding heavy blocking artifacts and thus requiring a postprocessing algorithm. Third, pure DL-based BCS pipelines \cite{8122281,8019428,8765626} have been proposed; unfortunately, existing frameworks typically use an undesirable fully connected network, and they are not fully recognized in theoretical analysis.

\subsection{Main contributions}
\label{sec:1.2}
To address the previously described problems, motivated by the perceptual learning archetype, we propose a pure DL-based framework for BCS from data acquisition to reconstruction in this paper. Figure~\ref{Fig:1} shows the schematic representation of our proposed AutoBCS architecture, which replaces the traditional BCS approach, with a unified image acquisition and reconstruction framework. Our main contributions are summarized as follows:
\begin{itemize}
  \item We build a bridge between traditional non-learning strategies and prior knowledge by training subblock image sets and developing a learning-based sensing matrix (LSM) for image acquisition without handcrafting the sparsifying domain and the sensing operator.
  \item The generated LSM is proven to satisfy the theoretical guarantees, such as the RIP, and it can thus be applied to the traditional BCS framework, while the sampling efficiency can be significantly improved.
  \item We develop a noniterative image reconstruction strategy that mainly relies on our customized octave reconstruction subnetwork, which establishes an end-to-end BCS reconstruction to eliminate blocking artifacts and maximize image reconstruction accuracy, in the proposed AutoBCS architecture.
\end{itemize}
The experimental results on several public testing databases all demonstrate that, compared with other state-of-the-art approaches, our AutoBCS framework can not only provide superior performance in terms of two image quality metrics (PSNR and SSIM) and visual perception but also automatically benefit reconstruction speed.

The rest of this paper is organized as follows. In Section \ref{sec:2}, we briefly review the related works to highlight our motivations. Section \ref{sec:3} describes our AutoBCS framework in detail. Extensive experiments are derived in Section \ref{sec:4} to demonstrate the superiority and effectiveness of the proposed framework. Section \ref{sec:5} presents the relevant discussions, and we conclude this paper in Section \ref{sec:6}.

\section{RELATED WORK}
\label{sec:2}
In this section, we will review some related works on both traditional block-based and DL-based image compressive sensing methods via general insights to highlight our motivations and contributions; we simultaneously introduce octave convolution.

\subsection{Traditional block-based image compressive sensing}
\label{sec:2.1}
Considering a full-size image $\bm {x}$ of $n= H \times W$ pixels, BCS divides $\bm{x}$ into nonoverlapping subblock images of size $A \times A$ and measures each subblock using the same sensing operator. Let $\bm{x}_i$ denote the vectorized vector of the $i^{th}$ subblock in a raster-scan manner. Using an $m_a \times A^2$ sensing matrix $\bm{B}_A$, we can obtain the corresponding measurements $\bm{y}_i$ via
\begin{equation}
\bm{y}_i = \bm{B}_A \bm{x}_i. \label{eq:4}
\end{equation}
Such a method is equivalent to applying general-purpose CS to the whole image $\bm {x}$ by using a block diagonal sensing operator $\bm{B}$, which is defined as:
 \begin{equation}
\bm{B}=\left(
  \begin{array}{cccc}
   \bm{B}_A       & 0          & \cdots      & 0 \\
    0             & \bm{B}_A   & \cdots      & 0 \\
    \vdots        & \vdots     & \ddots      & \vdots \\
    0             & 0          & \cdots      & \bm{B}_A \\
  \end{array}
\right).
\label{eq:5}
\end{equation}
Note that $m_a = \lfloor \frac {m A^2}{n}\rfloor$, and the sampling rate is $\tau =\frac {m}{n}$, where $m$ is the total number of measurements ($\bm{y}$) through $\bm{B}$.

In their inspiring work, Gan et al. \cite{gan2007block} used linear MMSE estimation as a criterion to obtain the initial solution and then employed two-stage iterative hard thresholding to further improve the quality of the initial image. Following this baseline framework, Mun et al. \cite{mun2009block} then developed an improved image reconstruction strategy for BCS, dubbed D-SPL, which combines SPL with direction transforms to simultaneously benefit smoothness and sparsity. This suggests that the dual-tree discrete wavelet transform (DWT) and contourlets can provide better reconstruction performances than other methods at low sampling rates. An improvement to this method that utilizes multiple scales and subbands of the DWT was proposed by Fowler et al. in \cite{fowler2011multiscale}. After that, they used a multi-hypothesis prediction to develop a different BCS strategy, dubbed MH-SPL, for images and videos \cite{chen2011compressed}. For still images, MH-SPL first calculates a multiple prediction for a subblock image from the spatially neighboring subblocks as an initial reconstruction and then obtains the final prediction for the subblock through an optimal linear combination. For video, MH-SPL can obtain the corresponding multi-hypothesis prediction of a frame from previously recovered adjacent frames. Comprehensive reviews on this kind of BCS method and its applications can be found in \cite{fowler2012block}. Moreover, block/patch-based methods that enforce the nonlocal self-similarity and local sparsity, such as GBsR \cite{zhang2014group} and WaPT \cite{zhao2019block}, were developed for natural images. These aforementioned approaches can always obtain high-quality images but with low reconstruction speeds.

In summary, traditional BCS strategies generally enhance performance by designing the sparsifying domain or exploiting the prior knowledge of the given image. Different from these approaches, our method is a pure data-driven BCS framework from image acquisition to reconstruction. Compared with the traditional BCS paradigm, our data-driven method automatically learns the features of each subblock image and the relationships among subblocks, and then it develops a learning-based sensing matrix from the training data to perform image acquisition. In addition, the proposed AutoBCS can provide better image recovery accuracy than other approaches, with an extremely fast recovery speed.

\begin{figure*}[t]
\center
\includegraphics [width=0.88 \textwidth,height=0.40 \textwidth]{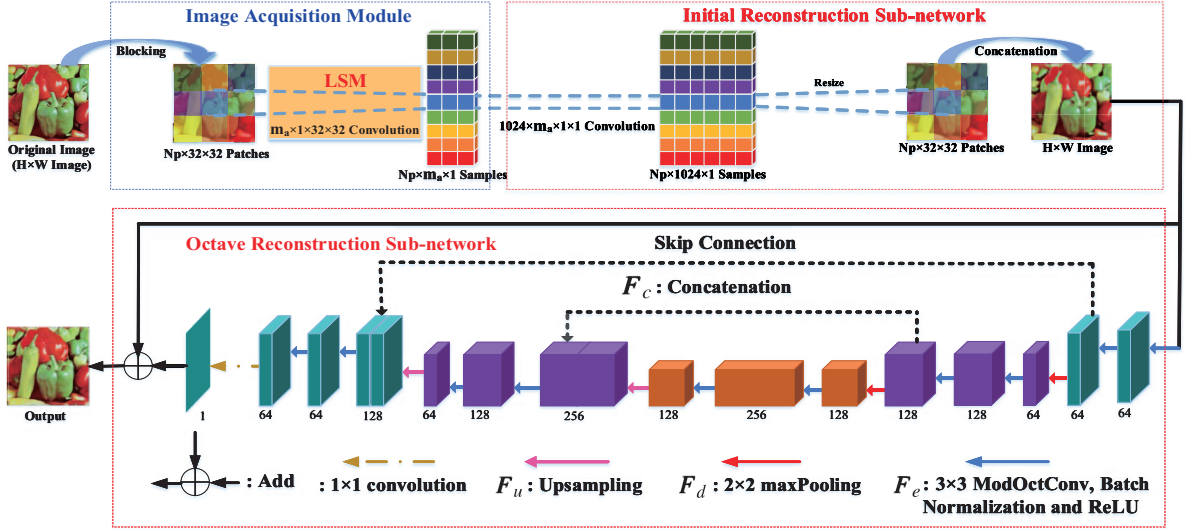}
\caption{The deep neural network architecture of AutoBCS contains two components: a data-driven image acquisition module and a noniterative data reconstruction module (composed of an initial reconstruction subnetwork and an octave reconstruction subnetwork). Note that we use different colors for the corresponding block processing operation to distinguish between different subblock images.}
\label{Fig:2}
\end{figure*}

\subsection{Deep learning-based image compressive sensing}
\label{sec:2.2}
To improve upon the traditional iterative reconstruction approaches, several attempts have been devoted to adopting the DL technique for CS imaging. Roughly speaking, we can divide these DL-based BCS frameworks into three categories. First, interpretable optimization-inspired image reconstruction strategies have been developed by casting iterative optimization algorithms into DL forms, in which the involved parameters are not hand-crafted but are gradually learned in an end-to-end manner. For example, Yang et al. \cite{yang2018admm} introduced an effective iterative deep architecture, called ADMM-CSnet, which was inspired by the iterative ADMM solver for solving Eq.~\eqref{eq:3}. Based on the ISTA solver, Jian et al. \cite{zhang2018ista} similarly developed ISTA-net to solve the proximal mapping problem related to the sparsity-inducing regularizer, and this indeed improved the recovery performance of their network. Nevertheless, these frameworks mainly focused on learning the sparsifying domain and reconstruction strategy without considering the sensing operator; thus, the data acquisition processes are not data-dependent, and this may weaken the sampling performance of such methods.

Second, block learning-based image reconstruction networks have been designed for CS imaging. Initially, Mousavi et al. \cite{mousavi2015deep} applied a stacked denoising autoencoder noniterative network (SDA-net) to reconstruct an image from its measurements. Following this work, Kulkarni et al. \cite{kulkarni2016reconnet} and Yao et al. \cite{yao2019dr2} developed two noniterative reconstruction frameworks, known as ReconNet and DR$^2$-net, respectively, by borrowing insights from traditional BCS reconstruction methods, which have competitive reconstruction performances with very short recovery speeds. However, such noniterative reconstruction networks usually have two deficiencies. First, these noniterative networks only employ intra-block information to recover a subblock image, thereby yielding blocking artifacts and thus requiring an additional deblocking algorithm with high computational complexity. In addition, the sensing matrix utilized for image acquisition is hand-crafted, i.e., these networks still do not take the sampling patterns of data acquisition into account.

To address these problems, pure DL-based BCS pipelines \cite{8122281,8019428,8765626} have been proposed for CS imaging, and these methods train a noniterative reconstruction network associated with learning the sampling patterns. For example, Wu et al. \cite{8019428} introduced a BCS framework called CS-net for jointly optimizing the sampling patterns and the reconstruction strategy via a convolutional neural network, and this framework was further extended in \cite{8765626}. In their research, CS-net could provide substantial improvements over other algorithms in terms of reconstruction accuracy, and it did so with a fast running speed. However, the existing pure DL-based BCS frameworks still suffer from many disadvantages, such as weak theoretical or comprehensive analyses and the use of undesirable fully connected or repetitive networks, which may hinder their practical applications.

Fundamentally, our proposed AutoBCS is a pure DL-based BCS framework that utilizes networks to learn a mapping between a set of measurements obtained by the LSM and high-quality images. Although both our proposed AutoBCS approach and other pure DL-based methods have similar inspirations, they are different for the following reasons. On one hand, our framework not only learns the prior information of images but also develops an LSM for image acquisition. The generated LSM is proven to satisfy the theoretical guarantees of the sampling patterns. On the other hand, existing reconstruction networks generally use either fully connected or repetitive convolutional layers to accomplish image reconstruction, while our customized noniterative reconstruction module in AutoBCS goes beyond that to consistently boost accuracy for images, thereby reducing computational complexity and memory costs.

\subsection{Octave convolution}
In convolutional neural networks, we can consider the output feature maps of a convolution layer as mixtures of information at multiple spatial frequencies. Octave convolution, introduced by Chen et al. \cite{chen2019drop}, is a novel frequency decomposition of convolution operations that stores and processes mixed feature maps while decreasing spatial redundancy. As an alternative to vanilla convolutions, it is a kind of plug-and-play convolutional operator that can effectively reduce the resolution of low-frequency maps and enlarge the receptive field, thus saving both computation and storage costs. As it also achieves significant performance gains, octave convolution is perfectly suitable for a variety of backbone deep convolutional networks related to image and video processing tasks, such as Res2Net \cite{8821313} and stabilizing GANs \cite{DBLP12534}, without any adjustment of the backbone network architecture.

To augment the detail recovery and maximize the image reconstruction accuracy of this method, we specifically modify the original octave convolution and customize an octave-transposed convolution to make the octave idea more suitable for our noniterative reconstruction network in this study. The related details about modified octave convolution and octave-transposed convolution are introduced in Appendix A, Part {\uppercase\expandafter{\romannumeral1}} and Part {\uppercase\expandafter{\romannumeral2}}, respectively. Compared with general DL-based strategies, our customized octave reconstruction strategy does not contain fully connected or repetitive networks, nor does it require deblocking processing.

\section{PROPOSED METHOD}
\label{sec:3}
In this section, we develop a pure DL-based BCS framework with data-driven image acquisition and noniterative data reconstruction to solve the questions posed in Section \ref{sec:1}, that is, how to efficiently measure images without handcrafting both the sparsifying domain and the sensing operator for improved sampling efficiency and how to use a noniterative reconstruction network to reconstruct images for optimal recovery accuracy.
\subsection{Framework of AutoBCS}
The proposed AutoBCS architecture is a pure DL framework incorporating data-driven image acquisition and reconstruction modules, as described in Fig.~\ref{Fig:2}. In the data acquisition module, AutoBCS automatically captures the features of each block image and the relationships among subblock images, and it correspondingly develops a learning-based sensing matrix (LSM) from the training data. In the image reconstruction module, AutoBCS learns a reconstruction mapping between a set of measurements obtained by the LSM and high-quality images. As the mapping is made, a low-dimensional and mutual joint manifold of these two types of data is implicitly learned, yielding an extremely expressive representation that maximizes image reconstruction accuracy. Specifically, the noniterative reconstruction module used in AutoBCS includes two phases: an initial reconstruction subnetwork and an octave reconstruction subnetwork. The former aims to obtain an initial recovery image with the proper global structure, while the latter focuses on augmenting fine details and finally outputs a high-quality reconstructed image.

For data training, the image acquisition module and the reconstruction module in AutoBCS are jointly optimized, and they form an end-to-end network to maximize both sampling efficiency and recovery performance. For application implementations, the trained LSM is utilized to yield a set of measurements for each subblock, and it equivalently forms a block-diagonal matrix with a constrained structure for encoding the full-size image to perform image acquisition. Moreover, the trained noniterative reconstruction module is considered a decoder for accomplishing image reconstruction.

\subsection{Data-driven image acquisition module}
As described in Eq.~\eqref{eq:4}, the traditional BCS strategy employs $\bm{B}_A$ to obtain a set of measurements $\bm{y}_i$ from each subblock image $\bm{x}_i$. In our AutoBCS framework, a convolution layer with kernel size ($A$), stride ($A$) and zero bias is used to implement the linear nonoverlapping block-based sampling stage, as each row of $\bm{B}_A$ can be considered a filter. More specifically, the image acquisition process of AutoBCS is shown in the upper left part of Fig.~\ref{Fig:2}, where $N_p$ denotes the total number of subblock images, and the size of each filter used in the sampling layer is the same as that of the subblock image, i.e., $A \times A$\footnote{As done in other BCS frameworks, we employ a block of size $A=32$ in our work.}. At a sampling rate of $\tau=\frac {m}{n}$, the sensing matrix $\bm{B}_A$ has $m_a = \lfloor \frac {m A^2}{n}\rfloor$ rows for image acquisition. As a consequence, the convolution layer has $m_a$ filters of size $A \times A$ for obtaining $m_a$ measurements. For example, if $\tau=0.01$, there are $10$ filters of size $32 \times 32$ in the sampling layer.

To be more formal, let $\bm{F}_A$ represent $m_a$ filters of the sampling convolution layer. Mathematically, the nonoverlapping block-based image acquisition process can be formulated as a convolution operation:
\begin{equation}
\bm{y}_i = conv(\bm{F}_{A}, \bm{x}_i) = \bm{F}_A \ast \bm{x}_i, \label{eq:6}
\end{equation}
which corresponds to the sampling procedure of traditional BCS, which is described by Eq.~\eqref{eq:4}. When inputting a subblock image $\bm{x}_i$ into the sampling layer, the output $\bm{y}_i$ is a vector of size $m_a$ that can be considered the corresponding measurements of $\bm{x}_i$. In the AutoBCS architecture, the sampling layer automatically learns the sampling patterns from the training data, i.e., the weights of $\bm{F}_A$ are gradually optimized for improved data acquisition. Once the training process is completed, we can obtain the corresponding LSM, denoted by $\bm{P}_A$. The LSM can naturally capture the features of each subblock image and the relationships among subblocks, and it can thus guarantee that $\bm{y}_i$ inherits more structural characteristics of $\bm{x}_i$ than those inherited with other methods.

\begin{figure}[t]
\center
\includegraphics [width=0.485 \textwidth,height=0.27 \textwidth]{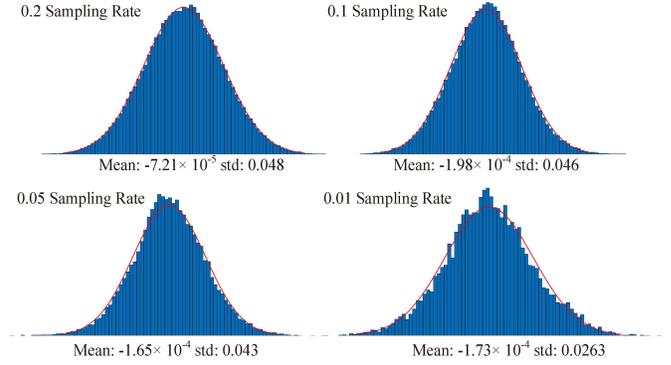}
\caption{Histograms of the element distribution of $\bm{P}_A$ in the cases: (a) $\tau=0.2$; (b) $\tau=0.1$; (c) $\tau=0.05$; (d) $\tau=0.01$, respectively, where the red line corresponds to standard Gaussian distribution. }
\label{Fig:3}
\end{figure}

To analyze the property of $\bm{P}_A$, we have trained four AutoBCS networks at four sampling rates, i.e., $\tau \in \{0.2, 0.1, 0.05, 0.01\}$. The training details will be introduced in Section~\ref{sec:5}. To this end, Fig.~\ref{Fig:3} plots the element distribution of the trained $\bm{P}_A$ for visualization purposes, where the red line corresponds to the standard Gaussian distribution. From these histograms, it can be distinctly observed that the elements of $\bm{P}_A$, ${P}_{{k,j}}$ $(1 \leq k \leq m_a, 1 \leq j \leq A^2)$ indeed obey Gaussian-like distributions, especially for AutoBCS with a high value of $\tau$.

Although the elements of the LSM and Gaussian matrix follow similar distributions, they are totally diverse in essence because of the following two aspects. First, the Gaussian matrix in the traditional framework is usually handcrafted without considering the prior information of the data, while our LSM is fundamentally learned according to prior knowledge rather than manually set. Second, the statistical properties of the elements of the Gaussian matrix, such as the mean and variance, are fixed and signal independent, while these same characteristics for the elements of the LSM are gradually optimized from the training data to improve the sampling efficiency of BCS. Roughly speaking, we can consider that $\bm{P}_A$ is automatically learned from the training data, and its elements follow a Gaussian-like distribution.

As described before, random matrices based on Gaussian or Bernoulli distributions or, more generally, a Gaussian-like distribution \cite{eldar2012compressed} have been shown to perfectly apply the theory of CS to satisfy real-world data acquisition demands. As a consequence, we can leverage the well-known theorems on random matrices to obtain many interesting conclusions regarding our trained LSM, and these conclusions come with theoretical guarantees. For brevity, we present the details in Appendix B. From another viewpoint, for a full-size image $\bm{x}$, the equivalent sensing operator $\bm{P}$ can be assumed to follow the form of Eq.~\eqref{eq:5} with constant block diagonal elements $\bm{P}_A$. As might be expected, the equivalent block sensing matrix $\bm{P}$ satisfies the RIP- or coherence-based performance guarantees as well. Additional related theoretical details on block diagonal matrices can be found in \cite{koep2019restricted} and the references therein.

In short, the trained LSM implicitly learns the structural characteristics of images and satisfies asymptotically optimal theoretical guarantees. As a result, the proposed AutoBCS approach is data-driven, and this automatically benefits image acquisition.

\subsection{Non-iterative data reconstruction module}
We implement our noniterative image reconstruction module with a deep neural network architecture consisting of an initial reconstruction subnetwork followed by an octave reconstruction subnetwork. The initial reconstruction subnetwork is essential since it can yield initial recovery images from the low-dimensional measurement vectors generated by the data acquisition module, and this is an indispensable preprocessing step for the octave reconstruction subnetwork to execute refinement because the proposed octave subnetwork only accepts images as its inputs.

\subsubsection{Initial reconstruction subnetwork}
Using linear MMSE estimation as the optimization criterion, traditional BCS frameworks \cite{gan2007block} utilize an $A^2 \times m_a$ reconstruction matrix $\widehat{\bm{B}}_A= \rho_{x_i,x_i} \bm{B}^{T}_A (\bm{B}_A \rho_{x_i,x_i} \bm{B}^{T}_A)^{-1}$ to obtain the initial reconstructed image, i.e., $\widehat{\bm{x}_i}=\bm{\widehat{B}}_A \bm{y}_i$, where $\widehat{\bm{x}_i}$ is the reconstructed vector of the $i^{th}$ subblock image and $\rho_{x_i,x_i}$ denotes the autocorrelation function for the input data. As a replacement for the reconstruction matrix, a convolutional layer with 1024 kernels of size $1 \times 1$ and no bias terms is used to perform the initial reconstruction process in our AutoBCS strategy. The initial reconstruction subnetwork is a linear signal reconstruction layer, as there is no activation function or bias.

Similar to the data acquisition procedure, we can also formulate the initial reconstruction process as a convolution operation:
\begin{equation}
\widehat{{\bm{x}}}_i = conv(\bm{F}_{int}, \bm{y}_i) =\bm{F}_{int} \ast \bm{y}_i, \label{eq:7}
\end{equation}
where $\widehat{{\bm{x}}}_i$ denotes the initial reconstruction vector of $\bm{x}_i$ and $\bm{F}_{int}$ represents $A^2$ filters of size $1 \times 1$ in the initial reconstruction layer. Obviously, $\widehat{{\bm{x}}}_i$ is a $1 \times A^2$ vector. The traditional BCS strategies always resize and concatenate these reconstruction vectors obtained by the reconstruction matrix to obtain the initial reconstruction result of $\bm{x}$. Following this baseline framework, we first resize each $1 \times A^2$ reconstruction vector $\widehat{{\bm{x}}}_i$ to an $A \times A$ subblock result and then concatenate all subblocks to obtain an initial reconstructed image $\widehat{{\bm{x}}}$ in our AutoBCS architecture. Let $\gamma(\cdot)$ and $\zeta(\cdot)$ denote the reshaping operator and the corresponding concatenation, respectively. We can formulate the following model:
 \begin{equation}
\widehat{{\bm{x}}}=\zeta\left(
  \begin{array}{cccc}
   \gamma (\widehat{{\bm{x}}}_{11})           & \gamma (\widehat{{\bm{x}}}_{12})            & \cdots      & \gamma (\widehat{{\bm{x}}}_{1c}) \\
   \gamma (\widehat{{\bm{x}}}_{21})           & \gamma (\widehat{{\bm{x}}}_{11})            & \cdots      & \gamma (\widehat{{\bm{x}}}_{2c}) \\
    \vdots        & \vdots     & \ddots       & \vdots \\
    \gamma (\widehat{{\bm{x}}}_{r1})          & \gamma (\widehat{{\bm{x}}}_{r2})            & \cdots      & \gamma (\widehat{{\bm{x}}}_{rc})   \\
  \end{array}
\right),
\label{eq:8}
\end{equation}
where $r$ and $c$ jointly denote the positions of the subblock images in the original full-size image.

In fact, only exploiting the initial reconstruction subnetwork is not sufficient because the subblock image quality of the initial recovery is low and the concatenation operator will always yield blocking artifacts in the space domain. For the sake of exact reconstruction, we develop a customized octave reconstruction subnetwork that can automatically draw on the information of both intra- and inter-subblock images to eliminate blocking artifacts and maximize image reconstruction accuracy.

\subsubsection{Octave reconstruction subnetwork}
As shown in the bottom part of Fig.~\ref{Fig:2}, an octave convolution-modified U-net architecture is introduced to further improve the reconstruction quality of $\bm{x}$. Similar to traditional U-net \cite{ronneberger2015u}, our proposed octave reconstruction subnetwork consists of a contracting path and an expanding path. The goal of the former is to produce high-dimensional features from the local receptive field. Thus, this part could be considered a feature extraction component that contains a set of octave-convolution and max-pooling operations (two convolutions followed by one max-pooling operation). For the contracting path, two separate operations $\bm{F}_e$ and $\bm{F}_d$ (denoted by the blue and red arrows in Fig.~\ref{Fig:2}, respectively) are involved, and they can be summarized as:
\begin{eqnarray}
\bm{F}_e(\widehat{{\bm{\chi}}})=ReLU(ModOctConv(\widehat{{\bm{\chi}}},3) + z), \\
\label{eq:9}
\bm{F}_d(\widehat{{\bm{\chi}}})=MaxPool(\widehat{{\bm{\chi}}},2),
\label{eq:10}
\end{eqnarray}
where $\widehat{{\bm{\chi}}}$ consists of the input feature maps, $ModOctConv(\cdot, 3)$ denotes the $3 \times 3$ modified octave convolution operation designed in Appendix A, Part {\uppercase\expandafter{\romannumeral1}}, $z$ is the bias of this layer, $ReLU(\cdot)$ represents one of the most common activation functions (rectified linear unit, $max(0, x)$), and $MaxPool(\cdot, 2)$ is the $2 \times 2$ max-pooling operation.

The upsampling path is designed to increase the resolution of the images to the original size, and more importantly, such a U-shaped design allows the network to perform convolutions on feature maps at different spatial resolutions, resulting in a multiscale feature representation with enlarged receptive fields. Thus, the first operation in this part is a transposed convolution for gradually increasing the resolution of the feature maps to the original size. This operation $\bm{F}_u$, denoted by the purple arrow in Fig.~\ref{Fig:2}, is included to oppose the effects of the pooling layers, and it can be expressed as:
\begin{eqnarray}
\bm{F}_u(\widehat{{\bm{\chi}}})=ReLU(OctTransConv(\widehat{{\bm{\chi}}},2) + z),
\label{eq:11}
\end{eqnarray}
where $OctTransConv(\cdot, 2)$ represents the $2 \times 2$ octave-transposed convolution introduced in Appendix A, Part {\uppercase\expandafter{\romannumeral2}}.

Immediately after these transposition operations, some concatenations between the layers of the contracting path and expanding path are introduced to compensate for the potential spatial loss incurred along the downsampling path. We can formulate the concatenation $\bm{F}_c$, denoted by the black dashed arrow in Fig.~\ref{Fig:2}, as:
\begin{eqnarray}
\bm{F}_c(\widehat{{\bm{\chi}}}_1, \widehat{{\bm{\chi}}}_2)=MatrixCat(\widehat{{\bm{\chi}}}_1, \widehat{{\bm{\chi}}}_2),
\label{eq:12}
\end{eqnarray}
where $\widehat{{\bm{\chi}}}_1$ and $\widehat{{\bm{\chi}}}_2$ denote the corresponding feature maps of the layers of the contracting path and expanding path, respectively. In other words, two concatenation layers $\bm{F}_c$ from the downsampling path to the upsampling path are used to compensate for the loss of spatial information by the max-pooling layers in the octave reconstruction subnetwork. For example, if the sizes of the matrices $\widehat{{\bm{\chi}}}_1$ and $\widehat{{\bm{\chi}}}_2$ are $256 \times 256 \times 10$ and $256 \times 256 \times 20$, respectively, then the output is a matrix of size $256 \times 256 \times 30$.

\begin{table*}[t]
\begin{center}
\vbox{\caption{AutoBCS vs. different conventional BCS methods on three typical benchmark databases}
\label{Tb:1}
\begin{tabular}{lcccccccccclcc}
\toprule
\hline
\multirow{2}{*}{Databases}                      & \multirow{2}{*}{\tabincell{c}{Sampling \\ rate $\tau$ } }     & \multicolumn{2}{c}{IRLS} & \multicolumn{2}{c}{D-SPL} & \multicolumn{2}{c}{MH-SPL} & \multicolumn{2}{c}{WaPT} & \multicolumn{2}{c}{GBsR} & \multicolumn{2}{c}{AutoBCS} \\ \cline{3-14}
   &          &     SSIM          &   PSNR \qquad   &    SSIM       &    PSNR  \qquad   &  SSIM         &   PSNR    \qquad  &  SSIM         &   PSNR  \qquad   &  SSIM         &     PSNR  \quad \quad   &      SSIM     &     PSNR     \\ \hline
\multicolumn{1}{l|}{\multirow{6}{*}{Set5}} & \multicolumn{1}{c|}{0.01} &0.4632 &15.87     &0.1387 &9.27     &0.4352 &18.01     &0.4398 &18.27    &0.4901 &18.78   &\textbf{0.6696} &\textbf{24.25}
\\
\multicolumn{1}{l|}{}                  & \multicolumn{1}{c|}{0.04}     &0.5145 &20.65     &0.2773 &12.71    &0.6101 &22.50     &0.6216 &22.81    &0.6888 &23.82   &\textbf{0.8446} &\textbf{29.18}
\\
\multicolumn{1}{l|}{}                  & \multicolumn{1}{c|}{0.1}      &0.7958 &27.26     &0.7641 &24.66    &0.8217 &28.63     &0.8435 &28.94    &0.8679 &30.12   &\textbf{0.9190} &\textbf{33.28}
\\
\multicolumn{1}{l|}{}                  & \multicolumn{1}{c|}{0.25}     &0.9023 &31.97     &0.8939 &32.64    &0.9055 &33.20     &0.9165 &33.97    &0.9401 &35.72   &\textbf{0.9607} &\textbf{37.65}
\\
\multicolumn{1}{l|}{}                  & \multicolumn{1}{c|}{0.3}      &0.9135 &33.02     &0.9062 &33.66    &0.9164 &34.08     &0.9247 &34.65    &0.9496 &36.89   &\textbf{0.9675} &\textbf{38.75}
 \\
\multicolumn{1}{l|}{}                  & \multicolumn{1}{c|}{Avg.}     &0.7179 &25.75     &0.5960 &22.58    &0.7378 &27.28     &0.7492 &27.73    &0.7873 &29.07   &\textbf{0.8723} &\textbf{32.62}
\\ \midrule
\multicolumn{1}{l|}{\multirow{6}{*}{Set14}} &\multicolumn{1}{c|}{0.01}  &0.4012 &15.51    &0.0978 &8.95    &0.4211 &17.22      &0.4254 &17.45    &0.4330 &17.81   &\textbf{0.5968}  &\textbf{23.12}
\\
\multicolumn{1}{l|}{}                  & \multicolumn{1}{c|}{0.04}      &0.5025 &19.04    &0.2170 &12.91   &0.5345 &21.02      &0.5598 &21.38    &0.5849 &22.32   &\textbf{0.7348}  &\textbf{28.67}
\\
\multicolumn{1}{l|}{}                  & \multicolumn{1}{c|}{0.1}       &0.7025 &25.88    &0.6908 &24.27   &0.7492 &26.88      &0.7583 &27.04    &0.7954 &28.24  &\textbf{0.8343}  &\textbf{29.56}
\\
\multicolumn{1}{l|}{}                  & \multicolumn{1}{c|}{0.25}      &0.8287 &29.45    &0.8264 &29.71   &0.8645 &31.20      &0.8785 &31.42    & 0.9017 &32.65  &\textbf{0.9205}  &\textbf{33.67}
\\
\multicolumn{1}{l|}{}                  & \multicolumn{1}{c|}{0.3}       &0.8574 &30.47    &0.8482 &30.68   &0.8824 &32.06      &0.8906 &32.43    &0.9092  &33.70  &\textbf{0.9347}  &\textbf{34.81}       \\
\multicolumn{1}{l|}{}                  & \multicolumn{1}{c|}{Avg.}      &0.6585 &24.07    &0.5360 &21.30   &0.6903 &25.68      &0.7025 &25.94    &0.7248  &26.94   &\textbf{0.8042}  &\textbf{29.97}
\\ \midrule
\multicolumn{1}{l|}{\multirow{6}{*}{BSD100}} &\multicolumn{1}{c|}{0.01} &0.4003  &16.07    &0.1065  &9.65     &0.4016 &18.06      &0.4197 &18.28    &0.4438 &18.64   &\textbf{0.5578} &\textbf{23.79}
\\
\multicolumn{1}{l|}{}                  & \multicolumn{1}{c|}{0.04}      &0.4539  &18.48    &0.2201  &13.30    &0.4954 &20.89      &0.5073 &21.12    &0.5390 &22.12   &\textbf{0.6833} &\textbf{26.40}
\\
\multicolumn{1}{l|}{}                  & \multicolumn{1}{c|}{0.1}       &0.6710  &25.53    &0.6341  &23.45    &0.6669 &25.17      &0.6728 &25.27    &0.7048 &25.88   &\textbf{0.7937} &\textbf{28.74}
\\
\multicolumn{1}{l|}{}                  & \multicolumn{1}{c|}{0.25}      &0.7902  &28.76    &0.7811  &28.25    &0.8026 &28.90      &0.8196 &29.58    &0.8432 &30.21   &\textbf{0.9019} &\textbf{32.33}      \\
\multicolumn{1}{l|}{}                  & \multicolumn{1}{c|}{0.3}       &0.8209  &29.55    &0.8086  &29.20    &0.8305 &29.85      &0.8434 &30.11    &0.8632 &30.90   &\textbf{0.9215} &\textbf{33.39}
\\
\multicolumn{1}{l|}{}                  & \multicolumn{1}{c|}{Avg.}      &0.6273  &23.68    &0.5101  &20.77    &0.6394 &24.57      &0.6526 &24.87    &0.6788 &25.55   &\textbf{0.7716} &\textbf{28.93}
\\ \hline
\bottomrule
\end{tabular}}
\end{center}
\end{table*}

Finally, a skip connection between the initial reconstruction result and the final output forms a residual block and can help the training process converge more easily. Therefore, the final reconstructed image $\tilde{\bm{x}}$ can be obtained by:
\begin{eqnarray}
\tilde{\bm{x}}= \widehat{{\bm{x}}} + OctReconNet(\widehat{{\bm{x}}}),
\label{eq:13}
\end{eqnarray}
where $OctReconNet(\widehat{{\bm{x}}})$ denotes the output of the octave reconstruction subnetwork.

\subsection{Loss function}
The proposed AutoBCS architecture is a pure DL-based BCS framework from image acquisition to reconstruction. Given an input image $\bm{x}$, our aim is to estimate the parameters $\bm{W}$ of AutoBCS such that it can automatically and efficiently obtain samples $\bm{y}$ via the image acquisition module and then rapidly and exactly reconstruct $\bm{x}$ from $\bm{y}$ through the data reconstruction module. To achieve this goal, we take the original image $\bm{x}$ as the ground truth and the corresponding reconstruction image $ {\tilde{\bm{x}}}$ generated by AutoBCS as the output. Specifically, we design two loss functions. One is used for training the whole AutoBCS framework, and the other is utilized for the initial reconstruction subnetwork. For the whole AutoBCS framework, the loss function is
\begin{eqnarray}
\mathcal{L}(\bm{W})= \frac {1}{2N} \sum_{j=1}^{N}\| AutoBCS({\bm{x}^j}; \bm{W}) - \bm{x}^j \|_2^2,
\label{eq:14}
\end{eqnarray}
where $N$ is the number of training images, $\bm{x}^j$ denotes the $j^{th}$ original image, and $AutoBCS({{{\bm{x}}^j}}; \bm{W})$ represents ${\tilde{\bm{x}}}^j$. For the initial reconstruction subnetwork, the loss function is
\begin{eqnarray}
\mathcal{L}_{int}(\bm{W}_{int})= \frac {1}{2N} \sum_{j=1}^{N}\| I({\bm{x}^j}; \bm{W}_{int}) - \bm{x}^j \|_2^2,
\label{eq:15}
\end{eqnarray}
where $\bm{W}_{int}$ and $I({\bm{x}^j}; \bm{W}_{int})$ represent the parameters and the output of the initial reconstruction subnetwork, respectively. Note that the image acquisition and reconstruction modules are jointly trained, which impels AutoBCS to overcome the two main challenges of CS theory.

\section{EXPERIMENTAL VALIDATION}
\label{sec:4}
The proposed AutoBCS is verified with extensive experiments in this section. In addition, we compare our AutoBCS with conventional BCS methods, including IRLS \cite{chen2018fast}, D-SPL \cite{mun2009block}, MH-SPL \cite{chen2011compressed}, WaPT \cite{zhao2019block}, and GBsR \cite{zhang2014group}, as well as newly developed DL approaches, including SDA-net \cite{mousavi2015deep}, ReconNet \cite{kulkarni2016reconnet}, DR$^2$-net \cite{yao2019dr2}, ISTA-net \cite{zhang2018ista}, and CS-net \cite{8765626}. Similar to that in our AutoBCS strategy, the block size for these state-of-the-art frameworks is set to $A=32$ as well. The structural similarity index (SSIM) and the peak signal-to-noise ratio (PSNR in dB) \cite{hore2010image} are used to evaluate the reconstruction accuracies of the network outputs.

\begin{figure}[tp]
\center
\includegraphics[width=0.485\textwidth,height=0.21\textwidth]{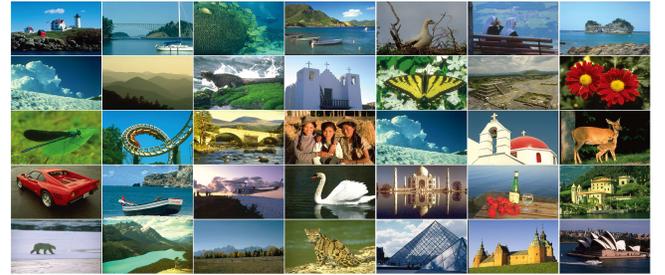}
\caption{Sample images from BSD500 database.}
\label{Fig:4}
\end{figure}

\begin{figure}[bp]
\center
\includegraphics [width=0.495 \textwidth,height=0.24 \textwidth]{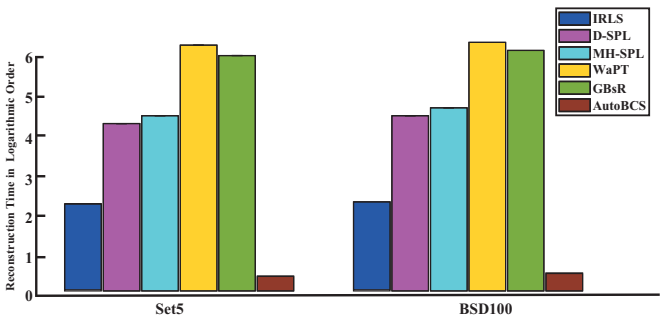}
\caption{Comparison of reconstruction time in logarithmic order for traditional BCS methods and AutoBCS on the Set5 and BSD100. Please zoom in for better comparison. }
\label{Fig:5}
\end{figure}

\begin{figure*}[t]
\center
\includegraphics [width=0.985 \textwidth,height=0.42 \textwidth]{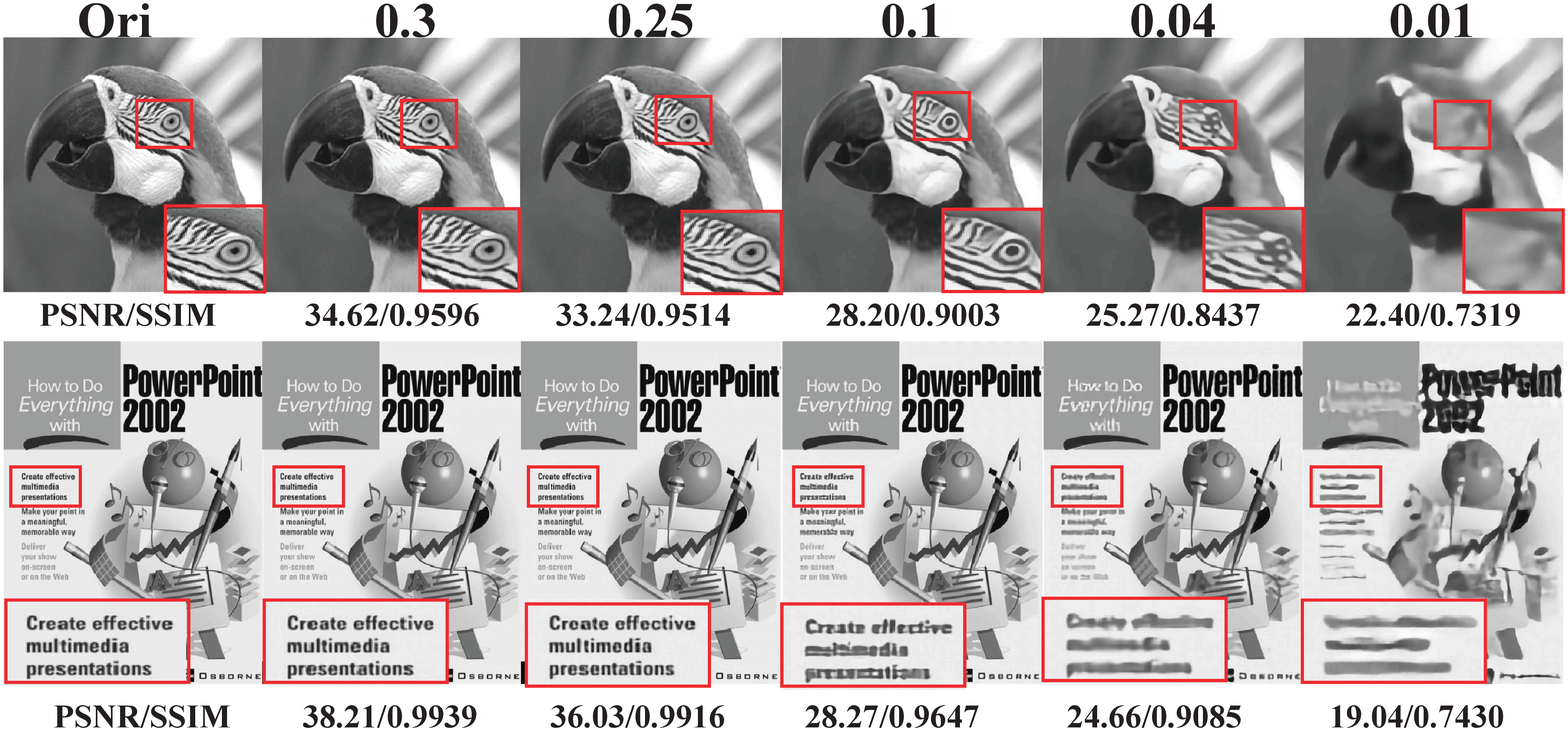}
\caption{Illustration of reconstructed images by using AutoBCS at different sampling rates. The first column: original image; the reconstruction columns for left to right correspond to $\tau=0.3$, $\tau=0.25$, $\tau=0.1$, $\tau=0.04$, and $\tau=0.01$, respectively. Please zoom in for better comparison.}
\label{Fig:6}
\end{figure*}

\subsection{Training details}
\subsubsection{Implementation}
The implementation of the proposed AutoBCS model was executed using the PyTorch 1.0 Framework on 2 TeslaV100 GPUs.

\subsubsection{Experimental data}
Our training dataset is composed of 400 images drawn from the popular BSD500 database (including 200 training images and 200 test images) \cite{amfmpami2011}. Some sample images used for training are illustrated in Fig.~\ref{Fig:4}. To improve network performance, data augmentation, including image flipping, rotating and their combination \cite{kim2016accurate}, is also applied to account for highly detailed recovery images cases. In total, the training images are prepared as 89600 sub-images ($96 \times 96$ pixel size), with a minibatch size of 64, for network training. We train the proposed AutoBCS model for 100 epochs, while the learning rate is scheduled as $10^{-3}$ for the first 50 epochs, $10^{-4}$ for epochs 51-80, and $10^{-5}$ for the final 20 epochs. In addition, five well-known datasets, Set5 \cite{bevilacqua2012low}, Set11 \cite{kulkarni2016reconnet}, Set14 \cite{zeyde2010single}, BSD68 and BSD100 \cite{martin2001d}, are utilized as test data to increase the diversity and generality of the benchmark. To facilitate visual perception, we only exploit the grayscale information of the images for both training and testing, as in other BCS approaches.

\subsection{Comparisons with traditional BCS methods}
In this subsection, the superiority and effectiveness of the proposed AutoBCS are validated by comparisons with five conventional BCS methods, namely, IRLS, D-SPL, MH-SPL, WaPT, and GBsR, on three typical benchmark databases, Set5, Set14 and BSD100. For these conventional algorithms, we use the default setups declared on the author's websites and run them on an Intel(R) Core(TV) i7-4770 CPU $@$ $3.40$ GHz with 16.0 GB of RAM. The comparison results in terms of the SSIM and PSNR for the aforementioned approaches are illustrated in Table~\ref{Tb:1}, and the best result for each unit is hereafter marked in boldface. From the overall view, it can be observed that our proposed AutoBCS model obtains the best performances in terms of both the SSIM and PSNR on the three databases compared to all compared conventional BCS methods. Even compared with the well-known classic algorithm GBsR, our AutoBCS model can enhance the SSIM by approximately $33.46\%$, $24.82\%$, $7.55\%$, $3.65\%$, and $3.74\%$ on average at different sampling rates, where $\tau \in \{0.01, 0.04, 0.1, 0.25, 0.3\}$, and the PSNR improvements are approximately 5.31 dB, 5.33 dB, 2.45 dB, 1.69 dB, and 1.82 dB, respectively. From these results, obviously, our proposed AutoBCS model has great performances at all sampling rates. Even when the sampling rate is 0.01, our method still provides superior performance in terms of both the SSIM and PSNR. Moreover, we plot a comparison of the recovery times of the competing algorithms on the Set5 and BSD100 databases for the case of $\tau =0.1$. Please refer to Fig.~\ref{Fig:5}, where the vertical axis follows $t=2+\log_{10}(t_{real})$ ($t_{real}$ represents the real recovery time). For example, if the real reconstruction time of AutoBCS $t_{real}=0.03$ ms, then $t=0.48$. Inspecting the figures, it can be concluded that our proposed AutoBCS model reduces the recovery time by a large amount and outperforms the other traditional BCS algorithms in all cases. Even compared with the fastest traditional BCS algorithm IRLS, AutoBCS is three orders of magnitude faster in terms of recovery speed.

\begin{table}[bp]
\begin{center}
\vbox{\caption{Average PSNR (dB) comparison between AutoBCS and various DL-based BCS methods on Set11}
\label{Tb:2}
\begin{tabular}{lcccccc}
\toprule
\hline
\multirow{2}{*}{Alg.} & \multicolumn{6}{c}{Sampling rate $\tau$} \\ \cline{2-7}
                                         & 0.01  \qquad    & 0.04 \qquad   & 0.1 \qquad      & 0.25 \qquad     & 0.3   \quad  \quad        & Avg.  \\ \hline
\multicolumn{1}{l|}{SDA-net}             & 17.29     & 20.12   & 22.65     &  25.34    & 26.63           &{22.41}    \\
\multicolumn{1}{l|}{ReconNet}            & 17.27     & 20.63   & 24.28     &  25.60    & 28.74           &{23.30}    \\
\multicolumn{1}{l|}{DR$^2$-net}          & 17.35     & 21.14   & 24.98     &  25.87    & 29.12           &{23.69}    \\
\multicolumn{1}{l|}{ISTA-net}            & 17.30     & 21.23   & 25.80     &  31.53    & 32.91           &{25.75}    \\
\multicolumn{1}{l|}{CS-net}              & 20.94     & 24.91   & 28.10     &  32.12    & 33.86           &{27.99}    \\
\multicolumn{1}{l|}{AutoBCS}             & \textbf{21.05} &\textbf{25.38}  &\textbf{28.78}  &\textbf{33.72} &\textbf{35.17} &\textbf{28.82}   \\ \hline
\bottomrule
\end{tabular}}
\end{center}
\end{table}

\begin{figure*}[tp]
\center
\includegraphics [width=0.985 \textwidth,height=0.50 \textwidth]{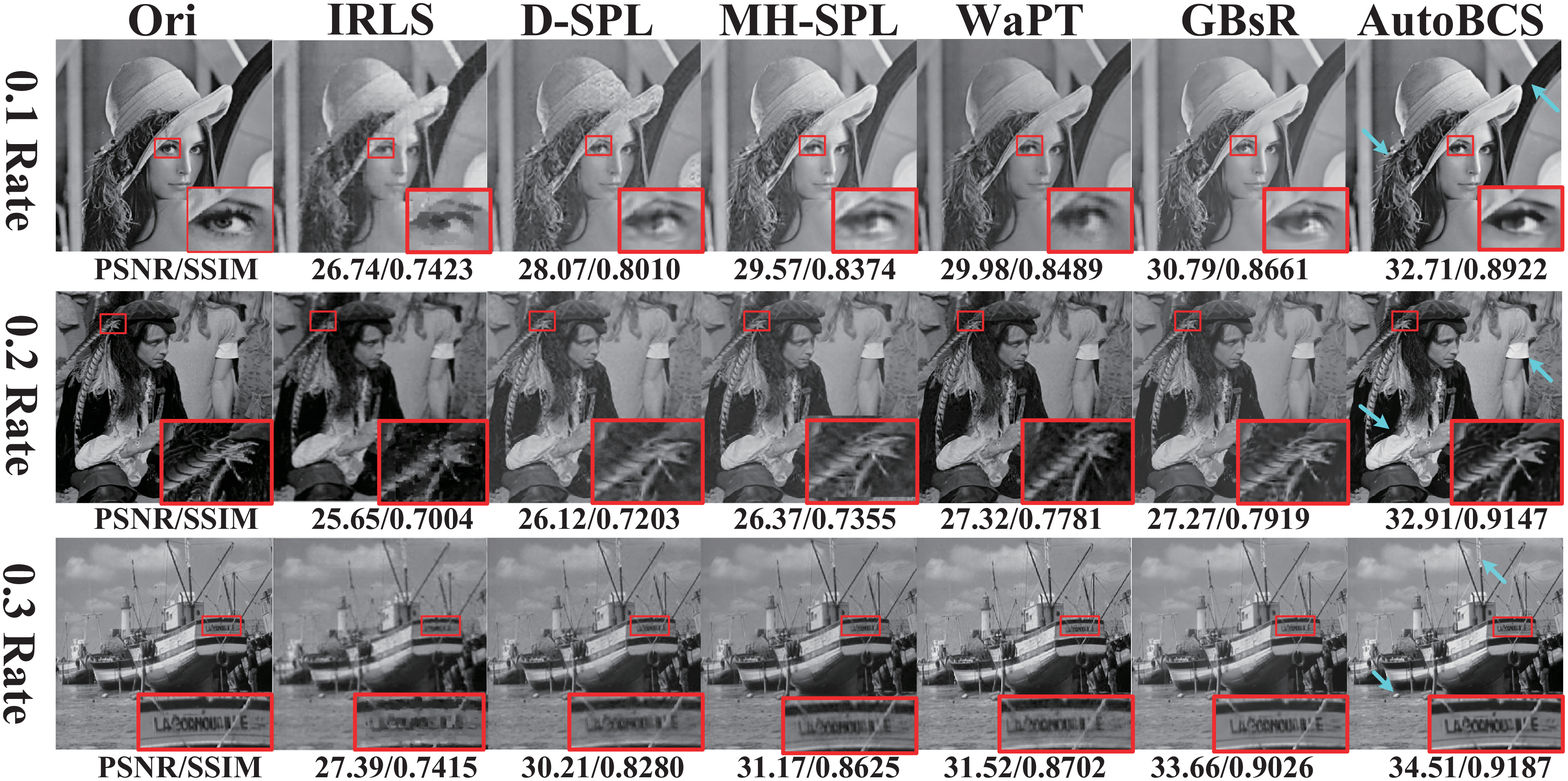}
\caption{Comparison for reconstruction images for traditional BCS methods and AutoBCS at various sampling rates. The first to third rows correspond to $\tau=0.1$, $\tau=0.2$, and $\tau=0.3$, respectively. Please zoom in for better comparison.}
\label{Fig:7}
\end{figure*}

In addition, we visualize the reconstructed images to facilitate a visual comparison. Figure~\ref{Fig:6} illustrates the corresponding reconstructed images obtained by AutoBCS with $\tau$ from 0.01 to 0.3. This once again verifies that AutoBCS can still capture rich semantic content, even at a particularly low sampling rate. As a consequence, our proposed AutoBCS model is extraordinarily desirable for low-power imaging devices. Finally, Fig.~\ref{Fig:7} depicts a comparison of the reconstructed images at various sampling rates. From these visual illustrations, we can clearly observe that AutoBCS is superior to the traditional BCS methods, especially with regard to recovering details and textures.

\begin{table}[t]
\begin{center}
\vbox{\caption{Average PSNR (dB) comparison between AutoBCS and various DL-based BCS methods on BSD68}
\label{Tb:3}
\begin{tabular}{lcccccc}
\toprule
\hline
\multirow{2}{*}{Alg.} & \multicolumn{6}{c}{Sampling rate $\tau$} \\ \cline{2-7}
                                         & 0.04  \qquad    & 0.1 \qquad   & 0.3 \qquad      & 0.4 \qquad     & 0.5   \quad  \quad        & Avg.  \\ \hline
\multicolumn{1}{l|}{SDA-net}             & 21.32     & 23.12   & 26.38     &  27.41    & 28.35           &  25.32       \\
\multicolumn{1}{l|}{ReconNet}            & 21.66     & 24.15   & 27.53     &  29.08    & 29.86           &  26.46       \\
\multicolumn{1}{l|}{DR$^2$-net}          & 21.87     & 24.34   & 27.77     &  29.27    & 29.96           &  26.64       \\
\multicolumn{1}{l|}{ISTA-net}            & 22.12     & 25.02   & 29.93     &  31.85    & 33.60           &  28.50       \\
\multicolumn{1}{l|}{CS-net}              & 24.03     & 27.10   & 31.45     &  32.53    & 34.89           &  30.00       \\
\multicolumn{1}{l|}{AutoBCS}             & \textbf{25.12}  &\textbf{27.46} &\textbf{32.18} &\textbf{34.23}  &\textbf{36.34}   &\textbf{31.07}       \\ \hline
\bottomrule
\end{tabular}}
\end{center}
\end{table}

In summary, our proposed AutoBCS model can improve upon the performances of other BCS methods in terms both reconstruction accuracy and recovery speed, and it can produce better results than those of the conventional BCS methods in different cases on these public benchmark datasets.

\subsection{Comparisons with DL-based BCS methods}

In this subsection, several DL-based BCS methods, namely, SDA-net, ReconNet, DR$^2$-net, ISTA-net, and CS-net, are compared with our AutoBCS strategy to further demonstrate the effectiveness of the proposed approach. Following previous work \cite{zhang2018ista}, we select the Set11 and BSD68 datasets for testing and summarize the PSNR results of these approaches for comparison purposes in Table~\ref{Tb:2} and Table~\ref{Tb:3}, respectively. For a fair comparison, the listed results are achieved from the corresponding works or reproduced as stated with the optimal settings provided by the corresponding papers. The denoising postprocessing operator of SDA-net, ReconNet and DR$^2$-net, i.e., the BM3D denoiser, is also considered, as it can significantly enhance the reconstruction accuracies of these methods. As illustrated in Table~\ref{Tb:2} and Table~\ref{Tb:3}, we can conclude that the proposed AutoBCS model outperforms all the competitive DL-based methods with these sampling rates.

\begin{table}[bp]
\begin{center}
\vbox{\caption{Model complexity of AutoBCS at different sampling rates}
\label{Tb:4}
\begin{tabular}{cccc}
\toprule
\hline
Sampling rate $\tau$   &  Input Resolution   & Params(M)   & GFLOPs    \\ \hline
0.01     & $256 \times 256$               & 1.81       & 20.10   \\
0.04     & $256 \times 256$               & 1.87       & 20.10    \\
0.1      &  $256 \times 256$              & 2.01       & 20.11    \\
0.25     &  $256 \times 256$              & 2.31       & 20.13    \\
0.3      &  $256 \times 256$              & 2.42       & 20.14    \\
Avg.     &  $256 \times 256$              & 2.08       & 20.12    \\ \hline
\bottomrule
\end{tabular}}
\end{center}
\end{table}

\begin{figure}[t]
\center
\includegraphics [width=0.495 \textwidth,height=0.25 \textwidth]{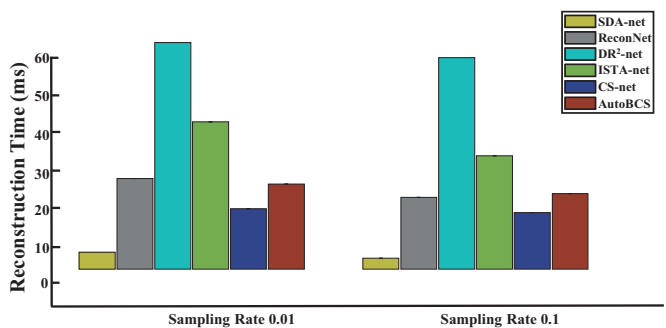}
\caption{Comparison of average reconstruction time (unit: ms) to reconstruct a $256 \times 256$ image for different DL-based BCS methods and AutoBCS on Set5 in case of $\tau =0.01$ and $\tau = 0.1$.}
\label{Fig:8}
\end{figure}

\begin{figure*}[tp]
\center
\includegraphics [width=0.985 \textwidth,height=0.35 \textwidth]{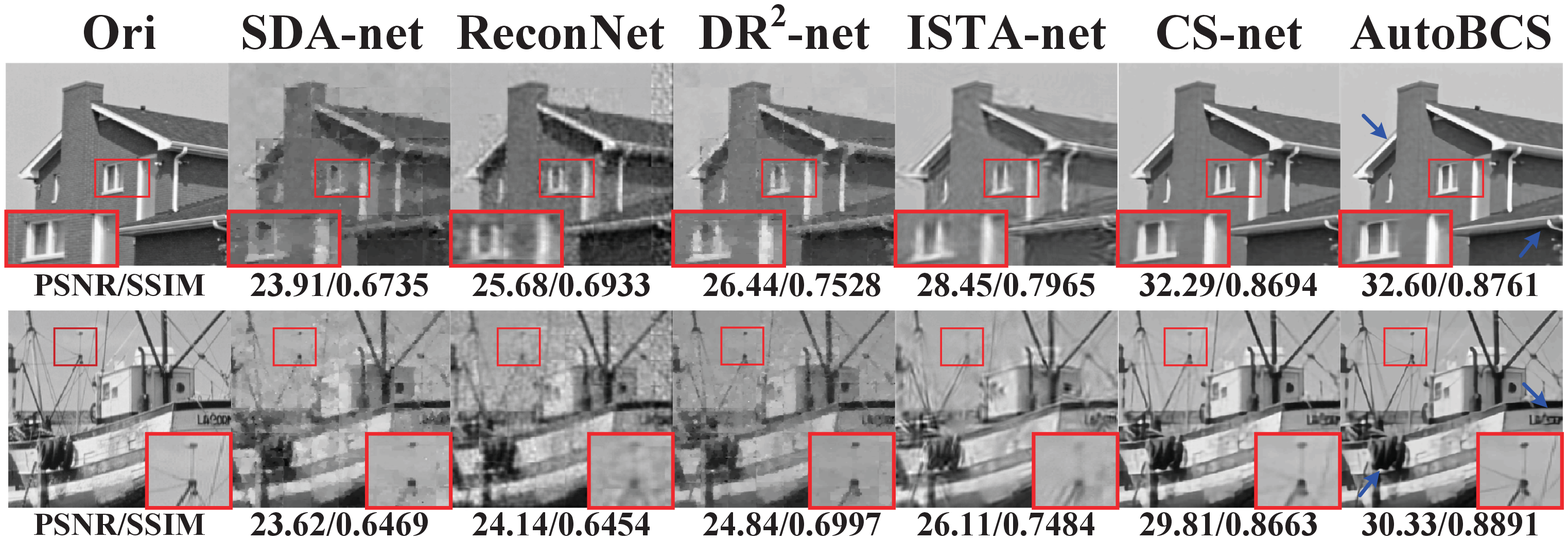}
\caption{Comparison of image reconstructions for various DL-based BCS approaches and AutoBCS at sampling rate $\tau=0.1$. The reconstruction columns from left to right correspond to SDA-net, ReconNet, DR$^2$-net, ISTA-net, CS-net, and AutoBCS, respectively. Please zoom in for better comparison.}
\label{Fig:9}
\end{figure*}

To carry out the model complexity analysis, we analyze our proposed AutoBCS from two aspects: $\#$Params and $\#$GFLOPs at various sampling rates, shown in Table~\ref{Tb:4}. The $\#$Params and $\#$GFLOPs are frequently used to weigh the space complexity and time complexity of the model, respectively. In aggregate, Table~\ref{Tb:4} shows that AutoBCS requires 20.12 GFLOPs in a single forward pass for a $256 \times 256$ pixel image. Moreover, the model complexity comparisons between AutoBCS and state-of-the-art DL-based frameworks including ReconNet, ISTA-net and CS-net are summarized in Table~\ref{Tb:5}. It should be noted that although the proposed AutoBCS contains the maximum number of parameters (2.01 million), it only needs lower GFLOPs than those of ISTA-net or CS-net to complete a single forward pass for a $256 \times 256$ pixel image at $\tau=0.1$. This contradiction in these models'space and time complexity should partly originate from the pooling-and-un-pooling design (or the contradicting-then-expanding design) and the OctConv operations in the reconstruction networks, which not only help enhance the multi-scale representation capability of the proposed model, but can also effectively reduce GFLOPs.

\begin{table}
\begin{center}
\vbox{\caption{Model complexity comparison between AutoBCS and state-of-the-art DL-based frameworks at $\tau=0.1$}
\label{Tb:5}
\begin{tabular}{lccc}
\toprule
\hline
Alg.   &  Input Resolution   & Params(M)   & GFLOPs    \\ \hline
ReconNet    & $256 \times 256$               & 0.32       & 1.92   \\
ISTA-net    & $256 \times 256$               & 0.34       & 34.55    \\
CS-net      &  $256 \times 256$              & 0.58       & 24.29    \\
AutoBCS     &  $256 \times 256$              & 2.01       & 20.11    \\ \hline
\bottomrule
\end{tabular}}
\end{center}
\end{table}

The reconstruction time of these approaches in the cases where $\tau =0.01$ and $\tau = 0.1$ are reported in Fig.~\ref{Fig:8}. We can evidentially conclude that, compared with traditional BCS approaches, DL-based BCS algorithms can achieve better performance, and this verifies the superiority of noniterative reconstruction methods. Relative to other aforementioned DL-based approaches, our proposed AutoBCS model obtains significantly better recovery accuracy with comparable reconstruction speed.

In particular, several visual comparisons between the reconstructed images for $\tau=0.1$ are illustrated in Fig.~\ref{Fig:9}. By checking the figures, we empirically observe that the reconstructed images output by SDA-net, ReconNet, and DR$^2$-net have heavy blocking artifacts. In contrast, our proposed AutoBCS model can smoothly reconstruct the image with more details and sharper edges. In summary, we can conclude from the results that AutoBCS significantly boosts its reconstruction performance over those of the other compared strategies on the aforementioned databases with respect to both quantitative validations and qualitative visualizations.

\begin{table*}[]
\begin{center}
\vbox{\caption{Comparison results for various sensing matrices with traditional BCS methods at $\tau=0.1$}
\label{Tb:6}
\begin{tabular}{clcccccccccccc}
\toprule
\hline
\multirow{2}{*}{Database}                      & \multirow{2}{*}{\tabincell{c}{Sensing \\ Matrix } }     & \multicolumn{2}{c}{IRLS} & \multicolumn{2}{c}{D-SPL} & \multicolumn{2}{c}{MH-SPL} & \multicolumn{2}{c}{WaPT} & \multicolumn{2}{c}{GBsR} & \multicolumn{2}{c}{Avg.} \\ \cline{3-14}
   &          &     SSIM          &   PSNR  \qquad  &    SSIM       &    PSNR  \qquad  &  SSIM         &   PSNR   \qquad   &  SSIM         &   PSNR    \qquad   &  SSIM         &     PSNR    \quad \quad  &      SSIM     &     PSNR     \\ \hline
\multicolumn{1}{c|}{\multirow{5}{*}{Set5}} & \multicolumn{1}{l|}{GSM} &0.7958 &27.26     &0.7641 &24.66      &0.8217 &28.63        &0.8435 &28.94        &0.8679 &30.12       &0.8186 &27.92            \\
\multicolumn{1}{c|}{}                  & \multicolumn{1}{l|}{BSM}     &0.7812 &27.11     &0.7747 &25.16      &0.8202 &28.56        &0.8453 &28.81        &0.8624 &29.86       &0.8168 &27.90
\\
\multicolumn{1}{c|}{}                  & \multicolumn{1}{l|}{BcSM}    &0.7967 &27.32     &0.7892 &26.08      &0.8238 &28.81        &0.8568 &29.07        &0.8701 &30.34       &0.8273 &28.32
\\
\multicolumn{1}{c|}{}                  & \multicolumn{1}{l|}{CbSM}    &0.7794 &27.18     &0.7571 &24.47      &0.8176 &28.40        &0.8431 &28.89        &0.8608 &29.73       &0.8116 &27.34
\\
\multicolumn{1}{c|}{}                  & \multicolumn{1}{l|}{LSM}  &\textbf{0.8463} &\textbf{29.87} &\textbf{0.8375} &\textbf{29.86} &\textbf{0.8499}  &\textbf{30.89}  &\textbf{0.8892} &\textbf{31.58}        &\textbf{0.9002} &\textbf{32.51}       &\textbf{0.8646} &\textbf{30.94}
\\ \hline
\bottomrule
\end{tabular}}
\end{center}
\end{table*}

\section{DISCUSSIONS}
\label{sec:5}
In this section, we first determine the LSM from a comparative perspective to show whether it can indeed enhance sampling efficiency, and then we discuss whether our proposed AutoBCS framework is robust to noise. Finally, we execute some ablation studies on our proposed AutoBCS architecture.

\subsection{LSM for traditional BCS frameworks}
\label{sec:5:1}

As previously described, we develop the LSM for data-driven image acquisition in our proposed AutoBCS architecture. The generated LSM automatically captures the feature of each block image and the relationships among subblock images, and it is proven to satisfy all theoretical guarantees. As a consequence, the LSM can significantly improve the sampling efficiency of our model and can be generally extended to traditional BCS frameworks. To quantify its performance, we apply the LSM and four other traditional sensing matrices (the GSM \cite{Candes1}, BSM \cite{eldar2012compressed}, BcSM \cite{gan2019chaotic} and CbSM \cite{gan2014compressive})\footnote{Note that GSM and BSM are two typical random sensing matrices, namely, Gaussian and Bernoulli sensing matrices. BcSM is a widely used deterministic binary code-based sensing matrix \cite{gan2019chaotic}. CbSM represents the Chebyshev chaotic binary matrix, which was introduced in \cite{gan2014compressive}. } in the aforementioned BCS algorithms, while all other parameters remain unchanged. In the case where $\tau=0.1$ and the Set5 database is chosen, the corresponding results w.r.t. both the SSIM and PSNR metrics are reported in Table~\ref{Tb:6}. According to the comparison results, we can distinctly observe that these traditional BCS methods have greatly enhanced sampling efficiency when equipped with our generated LSM for improved reconstruction, and the LSM can improve the reconstruction accuracies of these methods by roughly $5.62\%$ and $11.02\%$ in terms of the SSIM and PSNR on average, respectively.

\begin{figure}
\center
\includegraphics [width=0.495 \textwidth,height=0.28 \textwidth]{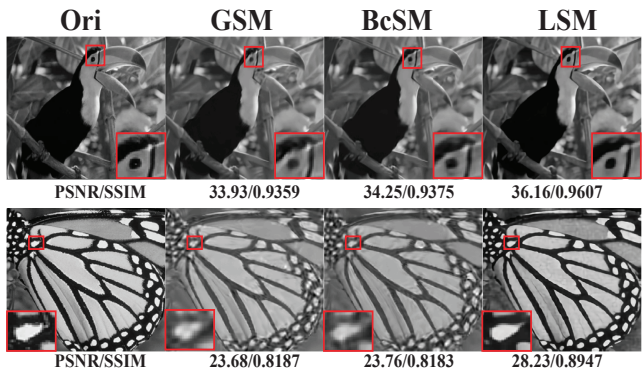}
\caption{Comparison of image reconstruction using three sensing matrices combined with traditional GBsR algorithm. First column: the original image; the reconstruction columns are based on GSM, BcSM and our trained LSM, respectively. }
\label{Fig:10}
\end{figure}

Let us now validate the comparison methods visually to make the distinction more convincing. We select ``bird'' and ``butterfly'' as the original images of interest. Following the previous settings, we present the comparison results in Fig.~\ref{Fig:10}. From these visual plots, we can distinctly observe that the images reconstructed with the LSM seem to be much clearer than those obtained using other matrices and contain richer textures with more edges and details. In total, the LSM can indeed improve recovery performance by a large margin over those of other classic sensing matrices.

\begin{table}[t]
\begin{center}
\vbox{\caption{Average PSNR (dB) comparison of the competing approaches for various noise levels on dataset Set5}
\label{Tb:7}
\begin{tabular}{clccccc}
\toprule
\hline
\multirow{2}{*}{Noise}         & \multirow{2}{*}{Alg.}      & \multicolumn{5}{c}{Sampling rate $\tau$} \\ \cline{3-7}
                                       &                                                  & 0.01      & 0.04    &  0.1      &  0.25     & 0.3          \\ \hline
\multicolumn{1}{l|}{\multirow{4}{*}{$\sigma_n =0.02$}}     & \multicolumn{1}{l|}{D-SPL}   & 9.26      & 12.71   & 24.81     & 32.39     & 33.21         \\
\multicolumn{1}{l|}{}                                      & \multicolumn{1}{l|}{MH-SPL}  & 18.15     & 22.59   & 28.51     & 32.77     & 33.60         \\
\multicolumn{1}{l|}{}                                      & \multicolumn{1}{l|}{GBsR}    & 18.42     & 23.05   & 29.21     & 34.11     & 35.13         \\
\multicolumn{1}{l|}{}                                      & \multicolumn{1}{l|}{AutoBCS} & \textbf{24.25}  & \textbf{29.11}  & \textbf{33.20}  & \textbf{37.15}    & \textbf{38.01}    \\ \hline
\multicolumn{1}{l|}{\multirow{4}{*}{$\sigma_n =0.05$}}     & \multicolumn{1}{l|}{D-SPL}   & 9.26      & 12.71   & 24.83     & 30.90     &  31.66        \\
\multicolumn{1}{l|}{}                                      & \multicolumn{1}{l|}{MH-SPL}  & 17.89     & 22.28   & 28.02     & 31.60     &  32.22       \\
\multicolumn{1}{l|}{}                                      & \multicolumn{1}{l|}{GBsR}    & 18.16     & 23.46   & 28.92     & 33.13     &  33.79       \\
\multicolumn{1}{l|}{}                                      & \multicolumn{1}{l|}{AutoBCS} & \textbf{24.24}   & \textbf{29.08}  & \textbf{32.73}   & \textbf{35.25}  & \textbf{35.49}     \\ \hline
\multicolumn{1}{l|}{\multirow{4}{*}{$\sigma_n =0.1$}}      & \multicolumn{1}{l|}{D-SPL}   & 9.26      & 12.71   & 24.26     & 28.83     & 29.18         \\
\multicolumn{1}{l|}{}                                      & \multicolumn{1}{l|}{MH-SPL}  & 17.30     & 20.96   & 26.41     & 29.34     & 29.64         \\
\multicolumn{1}{l|}{}                                      & \multicolumn{1}{l|}{GBsR}    & 18.25     & 21.68   & 26.66     & 30.07     & 30.37         \\
\multicolumn{1}{l|}{}                                      & \multicolumn{1}{l|}{AutoBCS} & \textbf{24.21}   & \textbf{28.77}  & \textbf{31.54}  & \textbf{31.81}   & \textbf{31.59}      \\ \hline
\bottomrule
\end{tabular}}
\end{center}
\end{table}

\begin{figure}[t]
\center
\includegraphics [width=0.495 \textwidth,height=0.54 \textwidth]{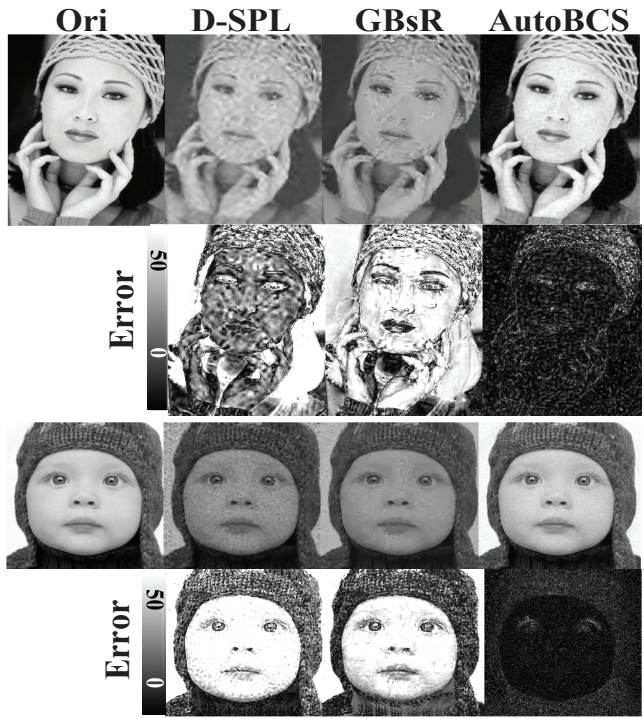}
\caption{Comparison of noisy image reconstruction for traditional BCS approaches (D-SPL and GBsR) and AutoBCS on Set5 in case of $\tau=0.1$ with Gaussian noise of $\sigma_n =0.1$, where the second and the fourth rows are corresponding image errors. Please zoom in for better comparison.}
\label{Fig:11}
\end{figure}

\subsection{AutoBCS for noisy reconstruction}
\label{sec:5:2}

This subsection aims to measure the reconstruction performance of AutoBCS in noisy situations\footnote{It is worth noting that here, we are still using the previously trained network instead of retraining a new AutoBCS under a noisy environment.}. To this end, three conventional BCS methods (D-SPL, MH-SPL, and GBsR) are simultaneously compared with AutoBCS in a setting with the Set5 dataset and additive Gaussian noise, where the standard deviations of the Gaussian noise are $\sigma_n =0.02$, $\sigma_n =0.05$, and $\sigma_n =0.1$, respectively. The average PSNR results obtained at various noise levels are shown in Table~\ref{Tb:7}. According to these results, all these methods, i.e., the proposed AutoBCS method and the traditional D-SPL, MH-SPL, and GBsR approaches, are more sensitive to noise at higher sampling rates. For example, as the noise level increases from $\sigma_n =0.02$ to $\sigma_n =0.1$ when the sampling rate $\tau=0.3$, the PSNR of the proposed AutoBCS method only decreases by 0.04 dB at $\sigma_n =0.02$; by contrast, a decrease of 6.39 dB is observed at $\sigma_n =0.1$.

Moreover, we can clearly see that our proposed AutoBCS model consistently outperforms the other competing approaches. For $\sigma_n =0.02$, our average PSNR gains over the D-SPL, MH-SPL, and GBsR methods are larger than 9.87 dB, 5.22 dB and 4.36 dB, respectively. For $\sigma_n =0.05$, AutoBCS can enhance the PSNR by approximately 9.49 dB, 4.96 dB and 3.87 dB on average, respectively. For $\sigma_n =0.1$, our proposed framework surpasses the D-SPL, MH-SPL, and GBsR methods by 8.74 dB, 4.85 dB and 4.18 dB, respectively. In particular, we illustrate the visual comparisons at a sampling rate of $\tau=0.1$ with Gaussian noise of $\sigma_n =0.1$ in Fig.~\ref{Fig:11}. From the visual plots, it can be observed that the other BCS algorithms all yield visually annoying artifacts; in contrast, the images recovered by the proposed AutoBCS method contain fewer artifacts and seem to be much smoother.

In summary, the proposed AutoBCS model achieves outstanding performances for noisy reconstruction according to the quantitative and visual results.

\subsection{Ablation study}
The influences of the octave reconstruction subnetwork, octave convolution, and generated LSM on the performance of the proposed AutoBCS framework are investigated in this subsection.

\begin{table}[bp]
\begin{center}
\vbox{\caption{Performance comparison between AutoBCS(U-net) and AutoBCS on three databases at various sampling rates.}
\label{Tb:8}
\begin{tabular}{llcccc}
\toprule
\hline
\multirow{2}{*}{\tabincell{c}{Sampling \\ rate}}   & \multirow{2}{*}{Databases} & \multicolumn{2}{c}{AutoBCS(U-net)} & \multicolumn{2}{c}{AutoBCS} \\\cline{3-6}
                             &                &   SSIM        &    PSNR        &    SSIM       &   PSNR       \\ \hline
\multicolumn{1}{l|}{\multirow{3}{*}{$\tau=0.04$}} & \multicolumn{1}{l|}{Set5}           &  0.8386       &    28.92       &  \textbf{0.8446}       &  \textbf{29.18}       \\
\multicolumn{1}{l|}{}                             & \multicolumn{1}{l|}{Set11}          &  0.7295       &    26.54       &  \textbf{0.7348}       &  \textbf{28.67}       \\
\multicolumn{1}{l|}{}                             & \multicolumn{1}{l|}{BSD100}         &  0.6786       &    26.26       &  \textbf{0.6833}       &  \textbf{26.40}       \\ \hline
\multicolumn{1}{l|}{\multirow{3}{*}{$\tau=0.1$}}  &  \multicolumn{1}{l|}{Set5}          &  0.9127       &    32.88       &  \textbf{0.9190}       &  \textbf{33.28}        \\
\multicolumn{1}{l|}{}                             &  \multicolumn{1}{l|}{Set11}         &  0.8285       &    29.36       &  \textbf{0.8343}       &  \textbf{29.56}        \\
\multicolumn{1}{l|}{}                             &  \multicolumn{1}{l|}{BSD100}        &  0.7877       &    28.57       &  \textbf{0.7937}       &  \textbf{28.74}         \\ \hline
\multicolumn{1}{l|}{\multirow{3}{*}{$\tau=0.25$}} &  \multicolumn{1}{l|}{Set5}          &  0.9503       &    36.46       &  \textbf{0.9607}       &  \textbf{37.65}        \\
\multicolumn{1}{l|}{}                             &  \multicolumn{1}{l|}{Set11}         &  0.9089       &    33.01       &  \textbf{0.9205}       &  \textbf{33.67}        \\
\multicolumn{1}{l|}{}                             &  \multicolumn{1}{l|}{BSD100}        &  0.8912       &    31.86       &  \textbf{0.9019}       &  \textbf{32.33}        \\ \hline
\bottomrule
\end{tabular}}
\end{center}
\end{table}

\begin{figure}
\center
\includegraphics [width=0.45 \textwidth,height=0.18 \textwidth]{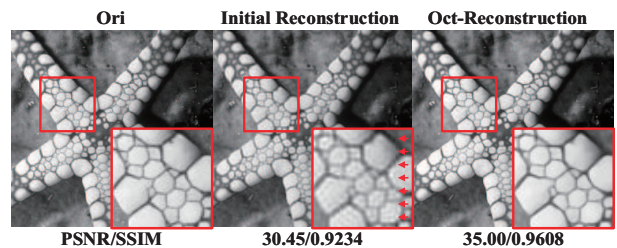}
\caption{Comparison of the results restored from the initial and octave reconstruction sub-networks of AutoBCS at $\tau=0.25$. Red arrows point to the blocking artifacts in the initial reconstruction output.}
\label{Fig:12}
\end{figure}

As depicted in the previous section, the initial reconstruction subnetwork is essential and indispensable in the proposed AutoBCS framework; however, the effectiveness of the octave reconstruction subnetwork has not yet been verified. Fig.~\ref{Fig:12} illustrates the output images from each subnetwork, showing that the octave reconstruction subnetwork can significantly alleviate artifacts and improve image quality. Overall, both reconstruction subnetworks are essential in the proposed AutoBCS scheme. Without the initial reconstruction subnetwork, $\bm{x}_i$ cannot be directly reconstructed by the octave reconstruction subnetwork from the measurement vector $\bm{y}_i$, while there would be heavy artifacts in the reconstruction results if there were no octave reconstruction subnetwork.

\begin{figure}
\center
\includegraphics [width=0.46 \textwidth,height=0.33 \textwidth]{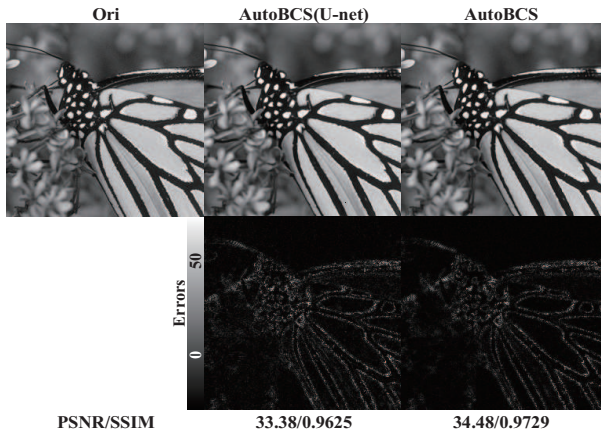}
\caption{Comparison of the reconstruction images for Auto-BCS(U-net) and the original AutoBCS at $\tau=0.25$.}
\label{Fig:13}
\end{figure}

To investigate the impact of the octave convolution on the reconstruction results, a U-net-based AutoBCS model, obtained by replacing the octave convolutions in Fig.~\ref{Fig:2} with vanilla convolutions and dubbed AutoBCS(U-net), is trained with the same training dataset. In other words, AutoBCS(U-net) and AutoBCS are completely the same except for the fine reconstruction subnetwork, where AutoBCS is equipped with an octave-based U-net. The qualitative and quantitative comparison results between AutoBCS(U-net) and AutoBCS are shown in Fig.~\ref{Fig:13} and Table~\ref{Tb:8}. These comparative results confirm that the octave convolution operation consistently improves the reconstruction results on different datasets at various sampling rates. Moreover, about $25\%$ relative increase in GFLOPs would be achieved for AutoBCS by replacing OctConv with the traditional convolutions in the backbone U-net. We refer the readers to the original OctConv work \cite{chen2019drop} for how and to what extent the OctConv can help reduce the model complexity.

\begin{figure*}[tp]
\center
\includegraphics [width=0.96 \textwidth,height=0.27 \textwidth]{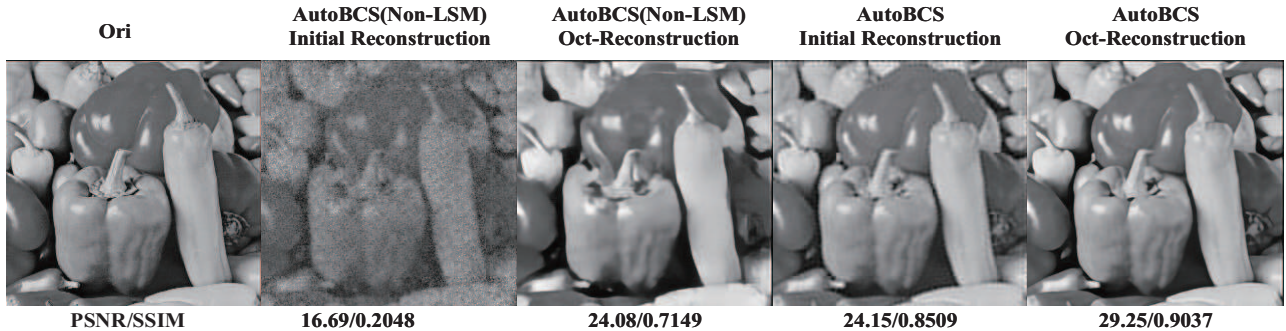}
\caption{Comparison of the reconstruction images from both the initial and octave reconstruction sub-networks for AutoBCS(Non-LSM) and AutoBCS at $\tau=0.1$. Please zoom in for better comparison.}
\label{Fig:14}
\end{figure*}

A comparative experiment is designed to elucidate the importance of the generated LSM for the proposed AutoBCS scheme. A variant of AutoBCS, called AutoBCS(Non-LSM), is obtained by using a fixed GSM instead of using the LSM, and it is trained with the same training datasets as before with the same training parameters. A comparison between the variant version and the original AutoBCS on an image at $\tau=0.1$ is shown in Fig.~\ref{Fig:14}. According to the results, the generated LSM can significantly enhance the quality of reconstructed images from both the initial and octave reconstruction subnetworks. In addition, the results illustrated in Fig.~\ref{Fig:10} and Table~\ref{Tb:5} confirm that the LSM can also improve the reconstruction performances of traditional BCS frameworks over those obtained with conventional sensing matrices, such as the GSM.

\section{CONCLUSIONS}
\label{sec:6}
In this work, we present a novel pure DL-based framework for BCS, called AutoBCS, which replaces the traditional BCS approaches from data acquisition to reconstruction. The proposed AutoBCS develops learning-based sensing patterns to perform data acquisition without handcrafting the sparsifying domain and the related parameters, and it customizes a subsequent inference model to accomplish fast image reconstruction with low computational costs. In summary, the generated LSM is data-driven, its sampling efficiency is theoretically guaranteed, and the computational complexity of the image reconstruction process is significantly reduced through our framework. Our proposed AutoBCS model is verified on several image databases, and the corresponding results demonstrate its effectiveness and superiority compared with traditional BCS approaches, as well as other newly developed DL-based methods.

One limitation of the proposed AutoBCS is that the sampling block size is set as a constant number in this study, which means that zero-padding and removing processing steps should be considered when the size of  original image is not divisible by this constant. A DL-based block compressed sensing with variational sampling matrix size for different sampling regions may be investigated in future work. Moreover, we will further investigate why the sampling efficiency of the LSM can outperform that of random matrices, even though we have shown that the LSM meets the theoretical requirements. Furthermore, some strong, robust modified DL strategies will be considered for noisy image reconstruction scenarios. Meanwhile, extending our proposed AutoBCS framework to other image inverse problems, such as inpainting and deconvolution, is another direction of concern for our future work.

\section*{Appendix A}
\subsection*{Part \uppercase\expandafter{\romannumeral1}: Modified octave convolution}
\begin{figure}[]
\center
\includegraphics [width=0.495\textwidth,height=0.22\textwidth]{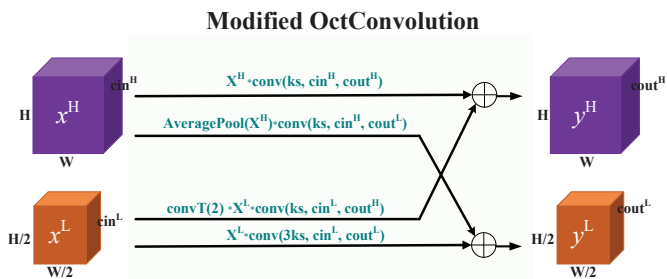}
\caption{The concept of modified octave convolution.}
\label{Fig:15}
\end{figure}

Let $\bm{X} \in \mathbb{R}^{ \{h \times w \times u \}}$ represent the input feature tensor of a convolution layer, where $h$, $w$, and $u$ denote the height, width, and channel of the feature maps, respectively. Unlike traditional vanilla convolutions, octave convolution explicitly factorizes the feature maps $\bm{X} $ along the channel dimension into $\{\bm{X}_H, \bm{X}_L\}$, where ${\bm{X}_H \in \mathbb{R}^{\{ h \times w \times (1-t)u \}}}$ denotes the high-resolution features, ${\bm{X}_L \in \mathbb{R}^{ \{ h \times w \times tu\}}}$ represents the low-resolution features, and $t \in [0,1]$. Typically, for the input layer of the octave network, $t=0$, and for the following layers, $t=0.5$. In our work, the original octave convolution is specifically modified by replacing the nearest interpolation with the transposed convolution. As a consequence, the corresponding outputs of the modified octave convolution, $\{\bm{Y}_H, \bm{Y}_L\}$, can be obtained as follows:
\begin{eqnarray}
\bm{Y}_H=conv(\bm{X}_H, \bm{W}_{H-H}) + convT(conv(\bm{X}_L, \bm{W}_{L-H}), 2), \nonumber \\
\bm{Y}_L=conv(\bm{X}_L, \bm{W}_{L-L}) + conv(pool(\bm{X}_H,2), \bm{W}_{H-L}), \nonumber
\end{eqnarray}
where $convT(\cdot,k)$ is the traditional transposed convolution operation with kernels of size $k \times k$, $pool(\bm{X},k)$ denotes the pooling operation with a kernel size of $k \times k$, and $\bm{W}$ contains the learnable weights of different layers. For visual observation purposes, we illustrate the concept of modified octave convolution in Fig.~\ref{Fig:15}.

Note that the replacement of the nearest interpolation by the transposed convolution results in the upsampling operation becoming a learnable block, and this block can be optimized during the training procedure of our network. Such a modification induces additional redundancy in the spatial dimension, and it leads to better multiscale representation learning than that achieved with vanilla convolutions.

\subsection*{Part \uppercase\expandafter{\romannumeral2}: Octave-transposed convolution}
An octave-transposed convolution is designed for our customized octave reconstruction subnetwork. Let $\{\bm{X}_H, \bm{X}_L\}$ be the input feature maps, and let $\{\bm{Y}_H, \bm{Y}_L\}$, $\bm{Y}_H \in \mathbb{R}^{ \{ 2h \times 2w \times (1-t)u \}}$, $\bm{Y}_L \in \mathbb{R}^{ \{ 2h \times 2w \times tu \}}$ represent the output features. Then, we can formulate the octave-transposed convolution as:
\begin{eqnarray}
\bm{Y}_H=convT(\bm{X}_H, \bm{W}_{H-H},2) + convT(\bm{X}_L,4), \nonumber\\
\bm{Y}_L=convT(\bm{X}_L, 2) + conv(\bm{X}_H, \bm{W}_{H-L}). \nonumber
\end{eqnarray}
Such an octave-transposed convolution allows the feature maps $\{\bm{X}_H, \bm{X}_L\}$ to double their spatial resolution and can help build the expanding part of the octave reconstruction subnetwork.

\section*{Appendix B}

First, we present the research on RIP-based guarantees for the LSM $\bm{P}_A$ by following the works of \cite{eldar2012compressed}.
\begin{thm}
\label{thm:1}
An $m_a \times A^2$ LSM, trained from the AutoBCS network, satisfies the $(s_a,\delta_a)$-RIP with the prescribed $\delta_a$ and any $s_a \leq c_0 A^2 / \log(m_a/s_a)$ with a probability exceeding $1-2e^{-c_1 m_a}$, where two constants $c_0, c_1 \geq 0$, and $s_a$ is the sparsity of an arbitrary signal $\in \mathbb{R}^{A^2}$.
\end{thm}

Theorem~\ref{thm:1} implies that the LSM achieves asymptotically optimal sampling performance. Moreover, this property straightforwardly allows us to pose other guarantees for the LSM, such as the coherence and spark property, via the Gershg\v{o}rin circle theorem. For example, the coherence property can be bridged to the $(s_a,\delta_a)$-RIP. Therefore, the following condition that guarantees sampling efficiency can be directly posed on our trained LSM.
\begin{thm}
\label{thm:2}
For an $m_a \times A^2$ LSM satisfying the $(s_a,\delta_a)$-RIP, the coherence $\mu(\bm{P}_A)$ and spark of the LSM follow $\mu(\bm{P}_A)= \delta_a / (s_a-1) $ and spark$(\bm{P}_A)>2s_a$, respectively.
\end{thm}
Very similar results in terms of sampling efficiency hold for random matrices, such as Gaussian and Bernoulli matrices. For each $s_a$-sparse signal to be uniquely represented by its samples $\bm{y}_s$, we can show that $\bm{P}_A$ meets the $(2s_a,\delta_a)$-RIP with $\delta_a > 0$, as this indicates that all sets of $2s_a$ columns of $\bm{P}_A$ are linearly independent, that is, spark$(\bm{P}_A)>2s_a$. From a theoretical perspective, the RIP of the LSM enables recovery guarantees that are more robust than those based on coherence and spark properties. In addition, we can draw similar conclusions by analyzing the elements' characteristics with respect to $\bm{P}_A$.

\bibliographystyle{ieeetr}
\bibliography{TMMRef0113}

\begin{thebibliography}{10}

\bibitem{Candes1}
E.~J. {Cand\`{e}s}, J.~{Romberg}, and T.~{Tao}, ``Robust uncertainty
  principles: Exact signal reconstruction from highly incomplete frequency
  information,'' {\em IEEE Transactions on Information Theory}, vol.~52, no.~2,
  pp.~489--509, 2006.

\bibitem{donoho2006compressed}
D.~L. Donoho, ``Compressed sensing,'' {\em IEEE Transactions on Information
  Theory}, vol.~52, no.~4, pp.~1289--1306, 2006.

\bibitem{Zhang8354810}
J.~{Zhang}, S.~{Cong}, Q.~{Ling}, and K.~{Li}, ``An efficient and fast quantum
  state estimator with sparse disturbance,'' {\em IEEE Transactions on
  Cybernetics}, vol.~49, no.~7, pp.~2546--2555, 2019.

\bibitem{Zhang7492261}
L.~Y. {Zhang}, K.~{Wong}, Y.~{Zhang}, and J.~{Zhou}, ``Bi-level protected
  compressive sampling,'' {\em IEEE Transactions on Multimedia}, vol.~18,
  no.~9, pp.~1720--1732, 2016.

\bibitem{Zhang8516287}
Y.~{Zhang}, Y.~{Xiang}, L.~Y. {Zhang}, Y.~{Rong}, and S.~{Guo}, ``Secure
  wireless communications based on compressive sensing: A survey,'' {\em IEEE
  Communications Surveys and Tutorials}, vol.~21, no.~2, pp.~1093--1111, 2019.

\bibitem{mei2017compressive}
G.~Mei, X.~Wu, Y.~Wang, M.~Hu, J.-A. Lu, and G.~Chen,
  ``Compressive-sensing-based structure identification for multilayer
  networks,'' {\em IEEE Transactions on Cybernetics}, vol.~48, no.~2,
  pp.~754--764, 2017.

\bibitem{Zhang8906045}
S.~{Zhang}, X.~{Li}, Q.~{Lin}, and K.~{Wong}, ``Nature-inspired compressed
  sensing for transcriptomic profiling from random composite measurements,''
  {\em IEEE Transactions on Cybernetics}, pp.~1--12, 2019.

\bibitem{eldar2012compressed}
Y.~C. Eldar and G.~Kutyniok, {\em Compressed sensing: Theory and applications}.
\newblock Cambridge University Press, 2012.

\bibitem{Liu6880351}
F.~{Liu}, L.~{Lin}, L.~{Jiao}, L.~{Li}, S.~{Yang}, B.~{Hou}, H.~{Ma},
  L.~{Yang}, and J.~{Xu}, ``Nonconvex compressed sensing by nature-inspired
  optimization algorithms,'' {\em IEEE Transactions on Cybernetics}, vol.~45,
  no.~5, pp.~1042--1053, 2015.

\bibitem{Zhou7900408}
Y.~{Zhou}, S.~{Kwong}, H.~{Guo}, X.~{Zhang}, and Q.~{Zhang}, ``A two-phase
  evolutionary approach for compressive sensing reconstruction,'' {\em IEEE
  Transactions on Cybernetics}, vol.~47, no.~9, pp.~2651--2663, 2017.

\bibitem{Lu7422756}
X.~{Lu}, Y.~{Yuan}, and X.~{Zheng}, ``Joint dictionary learning for
  multispectral change detection,'' {\em IEEE Transactions on Cybernetics},
  vol.~47, no.~4, pp.~884--897, 2017.

\bibitem{Wu7582406}
L.~{Wu}, Y.~{Wang}, and S.~{Pan}, ``Exploiting attribute correlations: A novel
  trace \protect{Lasso}-based weakly supervised dictionary learning method,''
  {\em IEEE Transactions on Cybernetics}, vol.~47, no.~12, pp.~4497--4508,
  2017.

\bibitem{Lin7875418}
L.~{Lin}, F.~{Liu}, L.~{Jiao}, S.~{Yang}, and H.~{Hao}, ``The overcomplete
  dictionary-based directional estimation model and nonconvex reconstruction
  methods,'' {\em IEEE Transactions on Cybernetics}, vol.~48, no.~3,
  pp.~1042--1053, 2018.

\bibitem{lai2016image}
Z.~Lai, X.~Qu, Y.~Liu, D.~Guo, J.~Ye, Z.~Zhan, and Z.~Chen, ``Image
  reconstruction of compressed sensing \protect{MRI} using graph-based
  redundant wavelet transform,'' {\em Medical Image Analysis}, vol.~27,
  pp.~93--104, 2016.

\bibitem{Baraniuk2010}
R.~G. Baraniuk, V.~Cevher, M.~F. Duarte, and C.~Hegde, ``Model-based
  compressive sensing,'' {\em IEEE Transactions on Information Theory},
  vol.~56, no.~4, pp.~1982--2001, 2010.

\bibitem{ravishankar2017low}
S.~Ravishankar, B.~E. Moore, R.~R. Nadakuditi, and J.~A. Fessler, ``Low-rank
  and adaptive sparse signal (\protect{LASSI}) models for highly accelerated
  dynamic imaging,'' {\em IEEE Transactions on Medical Imaging}, vol.~36,
  no.~5, pp.~1116--1128, 2017.

\bibitem{dong2014compressive}
W.~Dong, G.~Shi, X.~Li, Y.~Ma, and F.~Huang, ``Compressive sensing via nonlocal
  low-rank regularization,'' {\em IEEE Transactions on Image Processing},
  vol.~23, no.~8, pp.~3618--3632, 2014.

\bibitem{candes2008restricted}
E.~J. Cand\`{e}s, ``The restricted isometry property and its implications for
  compressed sensing,'' {\em Comptes Rendus Mathematique}, vol.~346, no.~9-10,
  pp.~589--592, 2008.

\bibitem{elad2010sparse}
M.~Elad, {\em Sparse and redundant representations: From theory to applications
  in signal and image processing}.
\newblock Springer Science \& Business Media, 2010.

\bibitem{cohen2009compressed}
A.~Cohen, W.~Dahmen, and R.~DeVore, ``Compressed sensing and best $k$-term
  approximation,'' {\em Journal of the American Mathematical Society}, vol.~22,
  no.~1, pp.~211--231, 2009.

\bibitem{li2014deterministic}
S.~Li and G.~Ge, ``Deterministic construction of sparse sensing matrices via
  finite geometry,'' {\em IEEE Transactions on Signal Processing}, vol.~62,
  no.~11, pp.~2850--2859, 2014.

\bibitem{gan2018construction}
H.~Gan, S.~Xiao, Y.~Zhao, and X.~Xue, ``Construction of efficient and
  structural chaotic sensing matrix for compressive sensing,'' {\em Signal
  Processing: Image Communication}, vol.~68, pp.~129--137, 2018.

\bibitem{gan2019chaotic}
H.~Gan, S.~Xiao, and F.~Liu, ``Chaotic binary sensing matrices,'' {\em
  International Journal of Bifurcation and Chaos}, vol.~29, no.~09, p.~1950121,
  2019.

\bibitem{fan2019compressed}
X.~Fan, Q.~Lian, and B.~Shi, ``Compressed sensing mri based on image
  decomposition model and group sparsity,'' {\em Magnetic Resonance Imaging},
  vol.~60, pp.~101--109, 2019.

\bibitem{zhang2020signal}
Y.~Zhang, X.~Li, G.~Zhao, B.~Lu, and C.~C. Cavalcante, ``Signal reconstruction
  of compressed sensing based on alternating direction method of multipliers,''
  {\em Circuits, Systems, and Signal Processing}, vol.~39, no.~1, pp.~307--323,
  2020.

\bibitem{chen2018fast}
C.~Chen, L.~He, H.~Li, and J.~Huang, ``Fast iteratively reweighted least
  squares algorithms for analysis-based sparse reconstruction,'' {\em Medical
  image analysis}, vol.~49, pp.~141--152, 2018.

\bibitem{monika2020adaptive}
R.~Monika, D.~Samiappan, and R.~Kumar, ``Adaptive block compressed sensing: A
  technological analysis and survey on challenges, innovation directions and
  applications,'' {\em Multimedia Tools and Applications}, pp.~1--18, 2020.

\bibitem{adcock2018infinite}
B.~Adcock, ``Infinite-dimensional compressed sensing and function
  interpolation,'' {\em Foundations of Computational Mathematics}, vol.~18,
  no.~3, pp.~661--701, 2018.

\bibitem{huynh2020fast}
T.~Huynh and R.~Saab, ``Fast binary embeddings and quantized compressed sensing
  with structured matrices,'' {\em Communications on Pure and Applied
  Mathematics}, vol.~73, no.~1, pp.~110--149, 2020.

\bibitem{cevher2009recovery}
V.~Cevher, P.~Indyk, C.~Hegde, and R.~G. Baraniuk, ``Recovery of clustered
  sparse signals from compressive measurements,'' tech. rep., Rice Univ Houston
  Tx Dept of Electrical and Computer Engineering, 2009.

\bibitem{polo2009compressive}
Y.~L. Polo, Y.~Wang, A.~Pandharipande, and G.~Leus, ``Compressive wide-band
  spectrum sensing,'' in {\em 2009 IEEE International Conference on Acoustics,
  Speech and Signal Processing}, pp.~2337--2340, IEEE, 2009.

\bibitem{ji2008multitask}
S.~Ji, D.~Dunson, and L.~Carin, ``Multitask compressive sensing,'' {\em IEEE
  Transactions on Signal Processing}, vol.~57, no.~1, pp.~92--106, 2008.

\bibitem{gan2007block}
L.~Gan, ``Block compressed sensing of natural images,'' in {\em 2007 15th
  International conference on digital signal processing}, pp.~403--406, IEEE,
  2007.

\bibitem{mun2009block}
S.~Mun and J.~E. Fowler, ``Block compressed sensing of images using directional
  transforms,'' in {\em 2009 16th IEEE International Conference on Image
  Processing}, pp.~3021--3024, IEEE, 2009.

\bibitem{fowler2011multiscale}
J.~E. Fowler, S.~Mun, and E.~W. Tramel, ``Multiscale block compressed sensing
  with smoothed projected landweber reconstruction,'' in {\em 2011 19th
  European Signal Processing Conference}, pp.~564--568, IEEE, 2011.

\bibitem{zhang2012image}
J.~Zhang, D.~Zhao, C.~Zhao, R.~Xiong, S.~Ma, and W.~Gao, ``Image compressive
  sensing recovery via collaborative sparsity,'' {\em IEEE Journal on Emerging
  and Selected Topics in Circuits and Systems}, vol.~2, no.~3, pp.~380--391,
  2012.

\bibitem{zhang2014group}
J.~Zhang, D.~Zhao, and W.~Gao, ``Group-based sparse representation for image
  restoration,'' {\em IEEE Transactions on Image Processing}, vol.~23, no.~8,
  pp.~3336--3351, 2014.

\bibitem{chen2011compressed}
C.~Chen, E.~W. Tramel, and J.~E. Fowler, ``Compressed-sensing recovery of
  images and video using multihypothesis predictions,'' in {\em 2011 Conference
  Record of the Forty Fifth Asilomar Conference on Signals, Systems and
  Computers}, pp.~1193--1198, IEEE, 2011.

\bibitem{zhao2019block}
H.~H. Zhao, P.~L. Rosin, and Y.~K. Lai, ``Block compressive sensing for solder
  joint images with wavelet packet thresholding,'' {\em IEEE Transactions on
  Components, Packaging and Manufacturing Technology}, vol.~9, no.~6,
  pp.~1190--1199, 2019.

\bibitem{yang2018admm}
Y.~{Yang}, J.~{Sun}, H.~{Li}, and Z.~{Xu}, ``\protect{ADMM-CSN}et: A deep
  learning approach for image compressive sensing,'' {\em IEEE Transactions on
  Pattern Analysis and Machine Intelligence}, vol.~42, no.~3, pp.~521--538,
  2020.

\bibitem{zhang2018ista}
J.~Zhang and B.~Ghanem, ``\protect{ISTA-N}et: Interpretable
  optimization-inspired deep network for image compressive sensing,'' in {\em
  Proceedings of the IEEE Conference on Computer Vision and Pattern
  Recognition}, pp.~1828--1837, 2018.

\bibitem{kulkarni2016reconnet}
K.~Kulkarni, S.~Lohit, P.~Turaga, R.~Kerviche, and A.~Ashok,
  ``Recon\protect{N}et: Non-iterative reconstruction of images from
  compressively sensed measurements,'' in {\em Proceedings of the IEEE
  Conference on Computer Vision and Pattern Recognition}, pp.~449--458, 2016.

\bibitem{yao2019dr2}
H.~Yao, F.~Dai, S.~Zhang, Y.~Zhang, Q.~Tian, and C.~Xu, ``\protect{DR}$^2$-net:
  Deep residual reconstruction network for image compressive sensing,'' {\em
  Neurocomputing}, vol.~359, pp.~483--493, 2019.

\bibitem{8122281}
A.~{Adler}, D.~{Boublil}, and M.~{Zibulevsky}, ``Block-based compressed sensing
  of images via deep learning,'' in {\em 2017 IEEE 19th International Workshop
  on Multimedia Signal Processing}, pp.~1--6, 2017.

\bibitem{8019428}
W.~{Shi}, F.~{Jiang}, S.~{Zhang}, and D.~{Zhao}, ``Deep networks for compressed
  image sensing,'' in {\em 2017 IEEE International Conference on Multimedia and
  Expo}, pp.~877--882, 2017.

\bibitem{8765626}
W.~Shi, F.~Jiang, S.~Liu, and D.~Zhao, ``Image compressed sensing using
  convolutional neural network,'' {\em IEEE Transactions on Image Processing},
  vol.~29, pp.~375--388, 2019.

\bibitem{fowler2012block}
J.~E. Fowler, S.~Mun, E.~W. Tramel, {\em et~al.}, ``Block-based compressed
  sensing of images and video,'' {\em Foundations and Trends in Signal
  Processing}, vol.~4, no.~4, pp.~297--416, 2012.

\bibitem{mousavi2015deep}
A.~Mousavi, A.~B. Patel, and R.~G. Baraniuk, ``A deep learning approach to
  structured signal recovery,'' in {\em 2015 53rd Annual Allerton Conference on
  Communication, Control, and Computing}, pp.~1336--1343, IEEE, 2015.

\bibitem{chen2019drop}
Y.~Chen, H.~Fan, B.~Xu, Z.~Yan, Y.~Kalantidis, M.~Rohrbach, S.~Yan, and
  J.~Feng, ``Drop an octave: Reducing spatial redundancy in convolutional
  neural networks with octave convolution,'' in {\em Proceedings of the IEEE
  International Conference on Computer Vision}, pp.~3435--3444, 2019.

\bibitem{8821313}
S.~{Gao}, M.~{Cheng}, K.~{Zhao}, X.~{Zhang}, M.~{Yang}, and P.~H.~S. {Torr},
  ``\protect{Res2Net}: A new multi-scale backbone architecture,'' {\em IEEE
  Transactions on Pattern Analysis and Machine Intelligence}, pp.~1--10, 2019.

\bibitem{DBLP12534}
R.~Durall, F.~Pfreundt, and J.~Keuper, ``Stabilizing \protect{GANs} with octave
  convolutions,'' {\em CoRR}, vol.~abs/1905.12534, 2019.

\bibitem{koep2019restricted}
N.~Koep, A.~Behboodi, and R.~Mathar, ``The restricted isometry property of
  block diagonal matrices for group-sparse signal recovery,'' {\em arXiv
  preprint arXiv:1901.06214}, 2019.

\bibitem{ronneberger2015u}
O.~Ronneberger, P.~Fischer, and T.~Brox, ``U-net: Convolutional networks for
  biomedical image segmentation,'' in {\em International Conference on Medical
  image computing and computer-assisted intervention}, pp.~234--241, Springer,
  2015.

\bibitem{hore2010image}
A.~Hore and D.~Ziou, ``Image quality metrics: \protect{PSNR} vs.
  \protect{SSIM},'' in {\em 2010 20th International Conference on Pattern
  Recognition}, pp.~2366--2369, IEEE, 2010.

\bibitem{amfmpami2011}
P.~Arbelaez, M.~Maire, C.~Fowlkes, and J.~Malik, ``Contour detection and
  hierarchical image segmentation,'' {\em IEEE Transactions on Pattern Analysis
  and Machine Intelligence}, vol.~33, no.~5, pp.~898--916, 2011.

\bibitem{kim2016accurate}
J.~Kim, J.~Kwon~Lee, and K.~Mu~Lee, ``Accurate image super-resolution using
  very deep convolutional networks,'' in {\em Proceedings of the IEEE
  Conference on Computer Vision and Pattern Recognition}, pp.~1646--1654, 2016.

\bibitem{bevilacqua2012low}
M.~Bevilacqua, A.~Roumy, C.~Guillemot, and M.~L. Alberi-Morel, ``Low-complexity
  single-image super-resolution based on nonnegative neighbor embedding,'' in
  {\em Proceeding of British Machine Vision Conference}, pp.~135.1--135.10,
  BMVA Press, 2012.

\bibitem{zeyde2010single}
R.~Zeyde, M.~Elad, and M.~Protter, ``On single image scale-up using
  sparse-representations,'' in {\em International Conference on Curves and
  Surfaces}, pp.~711--730, Springer, 2010.

\bibitem{martin2001d}
D.~Martin and C.~Fowlkes, ``A database of human segmented natural images and
  its application to evaluating segmentation algorithms and measuring
  ecological statistics,'' in {\em Proceedings of 8th International Conference
  on Computer Vision}, vol.~2, pp.~416--423, 2001.

\bibitem{gan2014compressive}
H.~Gan, Z.~Li, J.~Li, X.~Wang, and Z.~Cheng, ``Compressive sensing using
  chaotic sequence based on chebyshev map,'' {\em Nonlinear Dynamics}, vol.~78,
  no.~4, pp.~2429--2438, 2014.

\end{thebibliography}

\end{document}